\documentclass[preprint]{aastex}

\usepackage{natbib}  
\usepackage{graphicx}
\newcommand{\msun}{\ensuremath{M_{\odot}}}			
\newcommand{\lsun}{\ensuremath{L_{\odot}}}			
\newcommand{\hh}{\ensuremath{\textrm{H}_{2}}}			
\newcommand{\kms}{\textrm{km~s}\ensuremath{^{-1}}}	
\newcommand{\percc}{\ensuremath{\textrm{cm}^{-3}}}
\newcommand{\persc}{\ensuremath{\textrm{cm}^{-2}}}
\newcommand{\um}{\ensuremath{\mu m}}    
\newcommand{\htwo}{\ensuremath{\textrm{H}_2}}    
\newcommand{\HtwoO}{\ensuremath{\textrm{H}_2\textrm{O}}}    
\newcommand{\ha}{\ensuremath{\textrm{H}\alpha}}
\newcommand{\hb}{\ensuremath{\textrm{H}\beta}}
\newcommand{\necluster}{Sh~2-233IR~NE}
\newcommand{\swcluster}{Sh~2-233IR~SW}
\newcommand{\so}{ SO~\ensuremath{5_6-4_5} }
\newcommand{\ammonia}{NH\ensuremath{_3}}
\newcommand{\region}{IRAS 05358}
\newcommand{\twelveco}{\ensuremath{^{12}\textrm{CO}}}
\newcommand{\thirteenco}{\ensuremath{^{13}\textrm{CO}}}
\newcommand{\ceighteeno}{\ensuremath{\textrm{C}^{18}\textrm{O}}}
\def\ee#1{\ensuremath{\times10^{#1}}}
\newcommand{\degree}{\ensuremath{^{\circ}}}

\def\Figure#1#2#3#4{
\begin{figure}[htp]
\epsscale{#4}
\plotone{#1}
\caption{#2}
\label{#3}
\end{figure}
}

\def\Table#1#2#3#4#5#6{
\begin{deluxetable}{#1}
\tablewidth{0pt}
\tabletypesize{\scriptsize}
\tablecaption{#2}
\tablehead{#3}
\startdata
\label{#4}
#5
\enddata
#6
\end{deluxetable}
}

\shorttitle{Outflows in \region}
\shortauthors{Ginsburg, Bally, Yan, \& Williams}

\begin{document}

\title{Outflows and Massive Stars in the protocluster IRAS 05358+3543}

\author{Adam G. Ginsburg and John Bally}
\affil{Center for Astrophysics and Space Astronomy, Department of Astrophysical and Planetary Sciences}
\affil{University of Colorado}
\affil{389 UCB, Boulder, CO 80309-0389}
\email{Adam.Ginsburg@colorado.edu}
\email{John.Bally@colorado.edu}
\author{Chi-Hung Yan}
\affil{Institute of Astronomy and Astrophysics, Academia Sinica}
\affil{P.O. Box 23-141, Taipei, Taiwan}
\affil{Department of Earth Sciences, National Taiwan Normal University}
\affil{\#88 Sec. 4, Ting-chow Rd., Taipei, Taiwan}
\email{chyan@asiaa.sinica.edu.tw}
\author{Jonathan P. Williams}
\affil{Institute for Astronomy}
\affil{University of Hawaii}
\affil{2680 Woodlawn Dr. Honolulu, HI 96822}

\begin{abstract}
We present new near-IR \hh, CO J=2-1, and CO J = 3-2 observations to study
outflows in the massive star forming region IRAS 05358+3543. The
Canada-France-Hawaii Telescope \htwo\ images and James Clerk Maxwell
Telescope CO data cubes of the IRAS 05358 region reveal several new
outflows, most of which emerge from the dense cluster of sub-mm cores
associated with the \necluster\ cluster to the northeast of \region. We used
Apache Point Observatory (APO) JHK spectra to determine line of sight
velocities of the outflowing material. Analysis of archival VLA cm continuum
data and previously published VLBI observations reveal a massive star binary as
a probable source of one or two of the outflows.  We have identified probable
sources for 6 outflows and candidate counterflows for 7 out of a total of 11
seen to be originating from the \region\ clusters.  We classify the clumps
within \necluster\ as an early protocluster and \swcluster\ as a young
cluster, and conclude that the outflow energy injection rate approximately
matches the turbulent decay rate in \necluster.
\end{abstract}

\keywords{ISM: jets and outflows ---
          ISM: Herbig-Haro objects ---
          ISM: kinematics and dynamics ---
          ISM: individual: S233IR ---
          stars: formation}

\section{Introduction}

Collimated, bipolar outflows accompany the birth of young stars from the
earliest stages of star formation to the end of their accretion phase
\citep[e.g.][]{reipurth2001}.    While the birth of isolated low-mass stars is
becoming well understood, the formation of massive stars ($>10 \msun$) and clusters remains a
topic of intense study.    Observations show that moderate to high-mass stars
tend to form in dense clusters \citep{lada2003}.    In a clustered environment,  the dynamics of
the gas and stars can profoundly impact both accretion and mass-loss processes.
Feedback from these massive clusters may play a significant role in momentum
injection and turbulence driving in the interstellar medium.  

Outflows from massive stars are less studied than those from low mass stars
largely because massive stars accrete most of their mass while deeply embedded.
Therefore, unlike low mass young stars that are accessible in the optical,
massive stellar outflows can only be seen at infrared and longer wavelengths.
Direct evidence for jets from massive young stellar objects (YSOs) from \hh\ or
optical emission is generally lacking
\citep[e.g.][]{alvarez2005,kumar2002,wang2003}, although there is evidence that
massive stars are the sources of collimated molecular outflows from millimeter
observations \citep[e.g.][]{beuther2002b}.  Outflows from massive stars may
allow accretion to continue after their radiation pressure would
otherwise halt accretion in a spherically symmetric system
\citep{krumholz2009}.  They therefore represent a crucial component in
understanding how stars above $\sim$10 \msun\ can form.

\region\ is a double cluster of embedded infrared sources located at a distance
of 1.8 kpc in the Auriga molecular cloud complex \citep{heyer1996} associated
with the HII regions Sh-2 231 through 235 at Galactic coordinates around $l,~b$
= 173.48,+2.45 in the Perseus arm.   \necluster\ is the
collection of highly obscured and mm-bright sources slightly northeast of
\swcluster, which is the location of the IRAS 05358+3543 point source and the
optical emission nebula (see Figure \ref{fig:overview_ha}).  The IRAS source is
probably a blend of the three brightest infrared objects in the MSX A-band and
MIPS 24 \um\ images, which are located at \necluster, IR 41, and IR 6. For the
purpose of this paper,the whole complex including both sources is referred toas
\region, and otherwise refer to individual objects specifically.

Early observations revealed the presence of OH \citep{Wouterloot1993}, \HtwoO\ 
\citep{Scalise1989, Henning1992}, and methanol \citep{Menten1991} masers about
an arcminute northeast of the IRAS source, indicating that massive stars are 
likely present at that location.  Near infrared observations revealed
the presence of two embedded clusters  \citep{porras2000,jiang2001} labeled
\swcluster\ for the southwestern cluster associated with the IRAS source, and \necluster\ 
for the northeastern cluster located near the OH, \HtwoO, and CH$_3$OH masers.
Stars identified in \citet{porras2000} are referred to by the designation
``IR (number)'' corresponding to the catalog number in that paper.
\citet{porras2000} also included scanning Fabry-Perot velocity measurements of
the inner $\sim1$\arcmin.  CO observations revealed broad line wings indicative
of a molecular outflow \citep{casoli1986,shepherd1996}.  \citet{kumar2002} and
\citet{khanzadyan2004} presented narrow band images of 2.12 \um\ \htwo\
emission that reveled the presence of multiple outflows.  Interferometric
imaging of CO and SiO confirmed the presence of at least three flows emerging
from the northeast cluster centered on the masers \citep{beuther2002} having a
total mass of about 20 \msun .  \citet{beuther2002} also presented MAMBO 1.2 mm
maps and a mass estimate of 610 \msun\ for the whole region.
\citet{williams2004} presented SCUBA maps and mass estimates of the clusters of
195/126\msun\ for \necluster\ and 24/12 \msun\ for \swcluster\ (850 \um/450 \um).
\citet{Zinchenko1997} measured the dense gas properties using the \ammonia\
(1,1) and (2,2) lines.  They measure a mean density $n \approx 10^{3.60}$ \percc,
temperature 26.5K, and a mass of 600 \msun .  The total luminosity of the two
clusters is about 6300 \lsun , indicating that the region is giving birth to
massive stars \citep{porras2000}. 

Millimeter wavelength interferometry with arcsecond angular resolution has
revealed a compact cluster of deeply embedded sources centered on the \HtwoO\
and methanol maser position \citep{beuther2002,beuther2007,leurini2007}.
\citet{beuther2002} identified 3 mm continuum cores, labeled mm1-mm3 (shown in
Figure \ref{fig:outflowsh2}).  \citet{beuther2007} resolved these cores into
smaller objects.  Source mm1a is associated with a cm continuum point source
and will be discussed in detail below.

\region\ has previously been observed at low spatial resolution in the J=2-1 and J=3-2
transitions with the Kosma 3m telescope \citep{Mao2004}.  While the general presence
of outflows was recognized and a total mass estimated, the specific outflows were not 
resolved.  \citet{beuther2002} observed the CO J=6-5, J=2-1, and J=1-0
transitions at moderate resolution in the inner few arcminutes.
\citet{Thomas2008} observed C$^{17}$O in the J=2-1 and J=3-2 transitions with a
single pointing using the JCMT.

\section{Observations}

A collection of data acquired by the authors and from publicly
available archives is presented.  An overview of the data is presented in figure
\ref{fig:overview_ha}. The goal was to develop a complete picture of the outflows
in \region\ and their probable sources.  CO data were acquired to estimate the
total outflowing mass and to identify outflowing molecular material
unassociated with \hh\ shocks.  Archival Spitzer IRAC and MIPS 24 \um\ data
were used to identify probable YSOs as candidate outflow sources.
Near-infrared spectra were acquired primarily to determine \hh\ kinematics and
develop a 3D picture of the region.  Optical spectra were acquired to attempt
to identify stellar types in the unobscured \swcluster\ region.  Finally,
archival VLA data were used to acquire better constraints on the position and
physical properties of the known ultracompact HII (UCHII) region, and to detect
or set limits on other UCHIIs.

\subsection{Sub-millimeter Observations}

The 345 GHz J = 3-2 rotational transition of CO was observed with the James
Clerk Maxwell Telescope (JCMT) on 4 January,  2008 with the 16 element (14
functional) HARP-B heterodyne focal plane array.   Two  12\arcmin\ $\times$
10\arcmin\  raster scans in R.A.  and Dec.  were taken with orthogonal
orientations to assure complete coverage in the region of interest; this
resulted in a useable field 11.7\arcmin\ $\times$ 11.3\arcmin\ with higher noise
along the edges.  The beam size at 345 GHz is about 15\arcsec.

Observations were conducted during grade 3 conditions with the 225 GHz zenith
optical depth of the atmosphere $\tau\sim0.1$. A channel width of 488 kHz
corresponding to 0.423 \kms\ was used.    The maps required a total of 1 hour
to acquire and resulted in an effective integration time of 4.6 seconds per
pixel (there are 12,000 $6\times6\arcsec$\ pixels in the final grid), resulting
in a noise per pixel of 0.36 K \kms.

The optical depth and telescope efficiency corrections were applied by the JCMT
pipeline to convert the recorded antenna temperatures to the corrected antenna
\footnote{See \\
\url{http://docs.jach.hawaii.edu/JCMT/OVERVIEW/tel\_overview/} for a discussion
of JCMT parameters}.  An additional
main-beam correction has been applied, $$T_{mb}=\frac{T_A*}{\eta_{mb}}$$ where
$\eta_{mb} $ was measured by observing Mars to be $\approx0.60$ at 345
GHz.  Emission in the sidelobes is expected to be small at the outflow velocities.

On September 25 and November 15, 2008 the CO, $^{13}$CO, and C$^{18}$O J=2-1
transitions were observed in the central 3\arcmin\ of \region.  The beamsize
at 220 GHz is about 23\arcsec.  The sideband configuration used also includes
the \linebreak \nolinebreak{\so} and $^{13}$CS 5-4 transitions.  Conditions
during these observations were grade 5 ($\tau \sim 0.24-0.28$) and therefore
too poor to use the HARP instrument, but acceptable for the A3 detector.  

Data reduction used the Starlink package following the standard routines
recommended by the JCMT support scientists \footnote{
\url{http://www.jach.hawaii.edu/JCMT/spectral\_line/data\_reduction/acsisdr/}}.
The CO 3-2 data cube was extracted over a velocity range from --50 to 10 \kms\ LSR and
spectral baselines were fit over the velocity range --50 to --40 and 0 to 10
\kms\ and subtracted.  The data were re-gridded into  6\arcsec\  pixels  and 2
pixel Gaussian smoothing was used to fill in the gaps left by the two bad
detectors in the 4 $\times$4 array.   The data cube was cropped to remove
undersampled edges which have high noise and bad baselines.  The beam efficiency
was 0.68 at 230 GHz.

The A3 data cubes were extracted over the velocity range --60 to 20 \kms\ and
baselines were calculated over --60 to --40 and 0 to 20 \kms.  The data
was gridded into 10\arcsec\ pixels with 2 pixel gaussian smoothing to reduce
sub-resolution noise variations.

\subsection{Spitzer}

Spitzer IRAC bands 1 to 4 and MIPS band 1 data were retrieved from the Spitzer
Science Center archive.  \citet{qiu2008} acquired the data as part of a study
of many high-mass star forming regions; they identified YSO candidates based on
IRAC colors.  The version 18 post-BCD data products were used to produce images
and photometric catalog from \citet{qiu2008}, which was made from a more
carefully-reduced data set, was used for SED analysis.

\subsection{Near-IR images}
Near-infrared data were acquired using the Wide-field Infrared Camera (WIRCam) on
the Canada-France-Hawaii Telescope (CFHT) on Mauna Kea. The field of view is 
20\arcmin$\times$20\arcmin\ ~and pixel scale 0.3\arcsec.  Data were acquired on 
November 18, 19 and December 20, 2005.  The seeing was 0.5-0.7\arcsec\ during the
observations.  A 0.032 \um\ wide filter centered at 2.122 \um\ was used to take images
of the \hh\ S(1) 1-0 rovibrational transition.  Each \hh\ exposure was 58
seconds, and dithered images were taken for a total exposure time of 1755
seconds.  The data were reduced with the WIRCam pipeline.

\subsection{Near-IR spectra}
Near-infrared spectra were acquired using the TripleSpec instrument at Apache
Point Observatory.  TripleSpec simultaneously acquires J, H, and K band spectra
over a 42\arcsec\ long slit.  A slit width of 1.1\arcsec\ with an
approximate spectral resolution $\lambda/\Delta\lambda=2700$ was used.  

Observations were taken on the nights of December 2, 2008 and January 7, 2009.
Data on December 2 were taken in an ABBA nod pattern, but because of the need
to observe extended structure across the slit a stare strategy was selected on
January 7.  

The data were reduced using the {\sc twodspec} package in IRAF.  
HD31135, an A0 star, was used as a flux calibrator.  Wavelength calibration was
performed using night sky lines.  Lines filling the slit were subtracted
to remove atmospheric emission lines.  Telluric absorption correction was {\emph
not} performed, but telluric absorption is considered in the analysis.

The transformations from the observed geocentric reference frame to $v_{LSR}$ 
were computed to be 0.78 \kms\ on Dec 2 and 19.74 \kms\ on Jan 8.

\subsection{Optical Spectra}
Optical spectra were acquired using the Double Imaging Spectrograph instrument
at APO.  The high-resolution red and blue gratings
were centered at 6564 \AA\ and 5007 \AA\ with a coverage of about 1200 angstroms and
resolution $\lambda/\Delta\lambda \approx 5000$.  Sets of three 900s exposures
and three 200s exposures were acquired on the targets and on the spectrophotometric
calibrator G191-b2b with a 1.5" slit.  Observations were taken on the night
of January 17, 2009 under clear conditions.

Optical spectra were also reduced using the {\sc twodspec} package in IRAF.
Wavelength calibration was done with HeNeAr lamps and night sky lines
in the red band, and HeNeAr lamps in the blue band.  Lines filling the slit
were subtracted to remove atmospheric lines, though some astrophysical lines
also filled the slit and these were measured before background subtraction.
The $v_{LSR}$ correction for this date was 24.4 \kms.

\subsection{Optical imaging}

CCD images images were obtained on the nights of 14 and 15 September 2009
NOAO Mosaic 1 Camera at the f/3.1 prime focus 
of the 4 meter Mayal telescope atthe  Kitt Peak National Observatory (KPNO).  
The Mosaic 1 camera is a 8192$\times$8192 pixel array (consisting of eight 
2048$\times$4096 pixel CCD chips) with a pixel scale of 0.26$''$ pixel$^{-1}$ and 
a field of view 35.4$'$ on a side.  Narrow-band filters centered on 
6569\AA\ and 6730\AA\ both with a FWHM of 80\AA\ were use to obtain 
H$\alpha$ and [SII] images.  An SDSS i' filter which is centered on 
7732\AA\ with a FWHM of 1548\AA was used for continuum imaging.
A set of five dithered 600 second exposures were obtained in 
H$\alpha$ and [SII] using the standard MOSDITHER pattern 
to eliminate cosmic rays and the gaps between the
individual chips in Mosaic.  A dithered set of five 180 second 
exposures were obtained in the in the broad-band SDSS i-band filter to
discriminate between H$\alpha$, [SII], and continuum emission. 
Images were reduced in the standard manner by the NOAO Mosaic reduction
pipeline \citep{valdes2007}.

\Figure{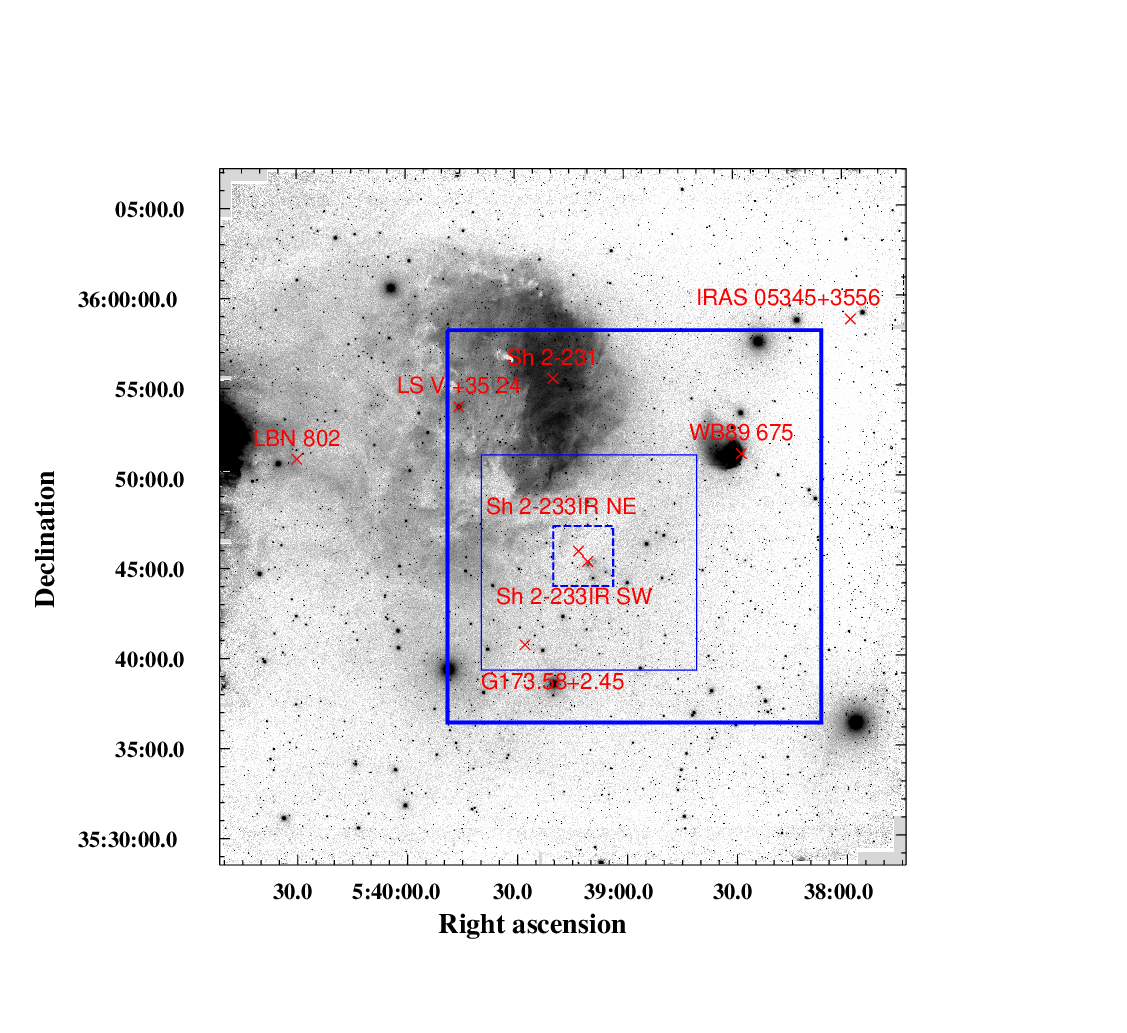}{The CFHT \hh\ (bold), CO 3-2 HARP
(thin), and CO 2-1 A3 (dashed) fields overlaid on the KPNO \ha\ mosaic
with selected objects identified by their SIMBAD names.  \swcluster\ is coincident
with IRAS 05358+3543. 
}{fig:overview_ha}{1.0}

\subsection{VLA data}
VLA archival data from projects AR482, AR513, AS831, and AM697 were re-reduced
to perform a deeper search for UCHII regions and aquire more data points on the
known UCHII's SED.  Data from AR482 were previously published in
\citet{beuther2007}, the other data are unpublished.  The data were reduced
using the VLA pipeline in AIPS ({\sc vlarun}).  The observations used and
sensitivities and beam sizes achieved are listed in Tables \ref{tab:vlatimes}
and \ref{tab:vla}.  There appeared to be calibration errors in the AR482
observations (the phase calibrator was 2-3 times brighter than in all other
observations) and this data were therefore not used in the final analysis, but
it produced consistent pointing results.

\Table{cccccccc}
{VLA Observation Program Names, Dates, and Times}
{\colhead{VLA } &  \colhead{Observation} &  \colhead{Time } &  \colhead{Array} &
\colhead{Band} & \colhead{Fluxcal} &  \colhead{Phase cal} & \colhead{Phase cal } \\
Observation & Date &on&&&&& Percent \\
Name &&Source&&&&& Uncertainty \\}
{tab:vlatimes}
{
AR482 & August 2 2001    & 2580s & B & X &3c286 & 0555+398   & 22  \\
AR513 & June 21 2003     & 7770s & A & X &3c286 & 0555+398   & 0.8 \\
AS831 & February 26 2005 & 2640s & B & X &3c286 & 0555+398   & 0.7 \\
AS831 & August 5 2005    & 2660s & C & X &3c286 & 0555+398   & 0.3 \\
AS831 & May 11 2006      & 2610s & A & X &3c286 & 0555+398   & 3.0 \\
AL704 & August 7 2007    & 6423s & A & Q &3c273 & 0555+398   & 18  \\ 
AL704 & September 1 2007 & 6423s & A & Q &3c273 & 0555+398   & 13  \\ 
AM697 & November 26 2001 & 2880s & D & Q &3c286 & 0555+398   & 2.2 \\
AM697 & November 28 2001 & 1530s & D & K &3c286 & 0555+398   & 2.1 \\
AM697 & November 28 2001 & 1530s & D & U &3c286 & 0555+398   & 5.8 \\
}{}

\section{Results}

\subsection{Near Infrared Imaging: Outflows and Stars}
Eleven distinct outflows have been identified in \region\ in the images.
Outflows are identified from a combination of J=3-2 CO data, shock excited
\htwo\ emission, and published interferometric maps \citep{beuther2002}.
Suspected CO outflows were identified by the presence of wings on the CO J=3-2
emission lines that extended beyond the typical velocity range of emission
associated with the line core.    The single dish data were compared to the
interferometric maps of \citet{beuther2002}.  The CFHT \hh\ image was then used
to search for shock-excited emission associated with the outflow lobes.

\begin{figure*}[htpb]
    \center
    \epsscale{0.75}
    \hspace{-1.2in}
    \plotone{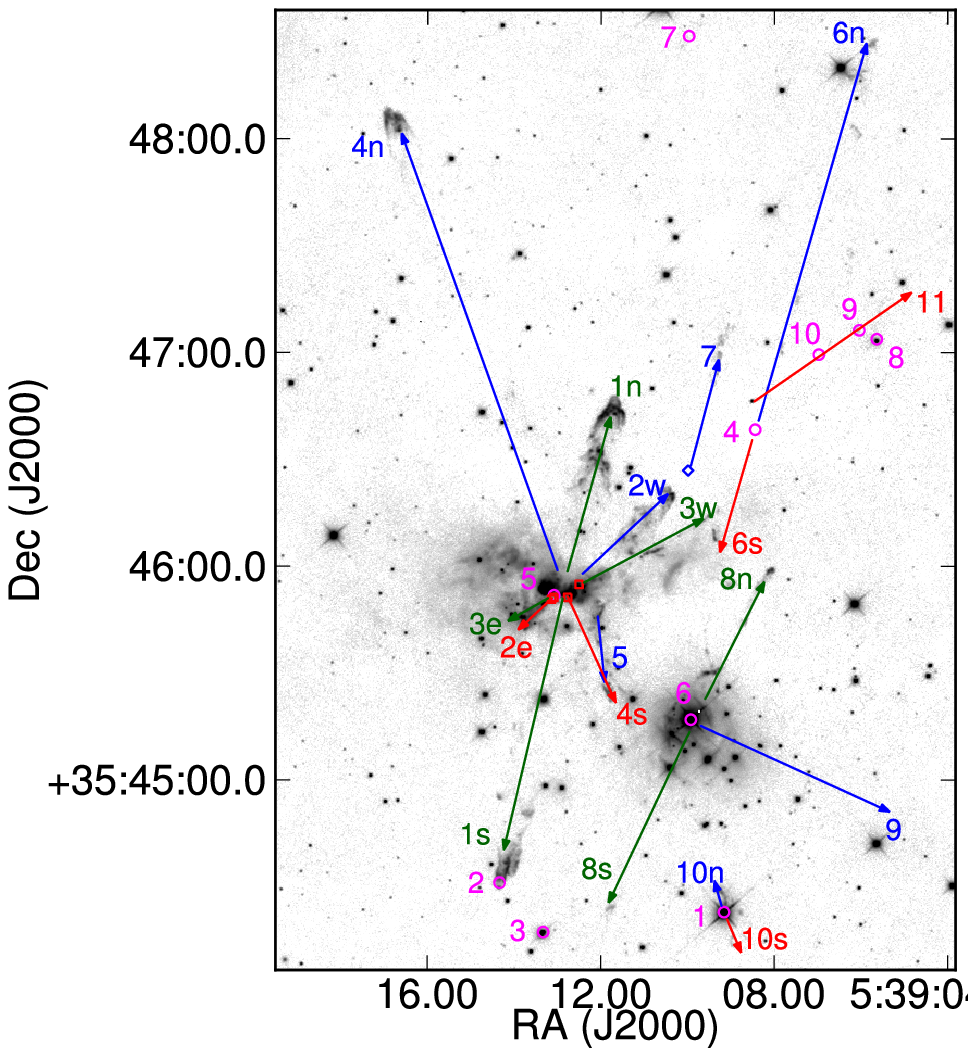}
    \caption{The outflows described in section \ref{sec:outflows} overlaid on
    the CFHT \hh\ image.  Numbers followed by {\it r} and {\it b} (red and
    blue), {\it n} and {\it s} (north and south), or {\it e} and {\it w} (east
    and west) are thought to be counterflows.  Red and blue vectors indicate
    red and blue doppler shifts.  Green vectors indicate where the doppler
    shift is ambiguous or cannot be determined.  Magenta circles are Spitzer
    24\um\ sources.  Red squares are \citet{beuther2002} mm sources (from left
    to right, mm1, mm2, mm3).  The blue diamond is a YSO candidate detected
    only in IRAC bands.  The length of the vectors corresponds to the
    approximate length of the outflows.  Source 1 and 6 correspond to
    \citet{porras2000} IR 6 and IR 41 respectively, and they are discussed under
    these names in sections \ref{sec:outflows}.  The bows of Outflow 1n and 4n 
    are detected in \ha\ and [S II] emission and are therefore as identified
    as Herbig-Haro objects HH 993 and 994 respectively.
    \label{fig:outflowsh2}}
\end{figure*}

\Figure{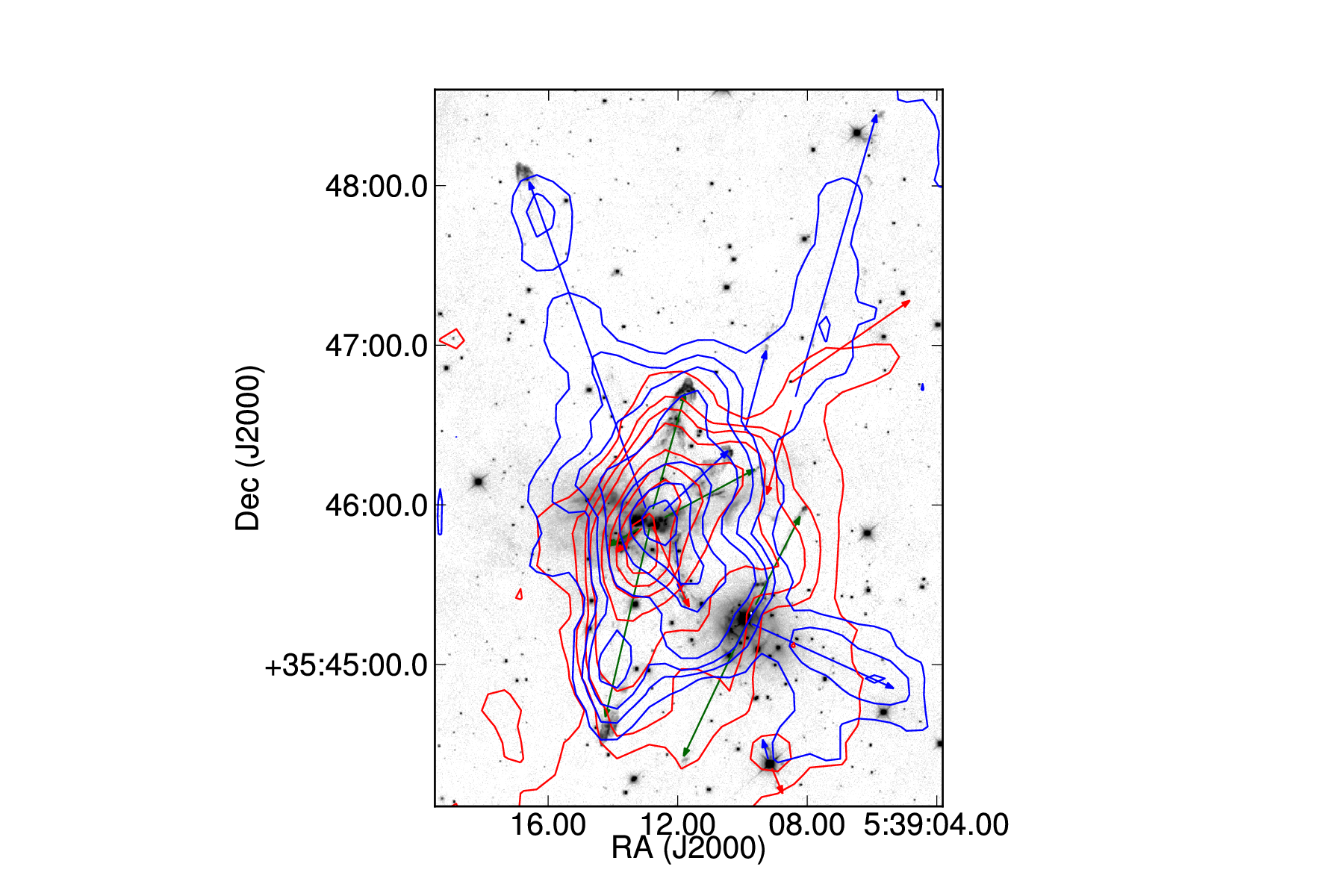} {CO contours integrated from $v_{LSR}=$ -13 to -4
\kms\ (red) and -34 to -21 \kms\ (blue) at levels of 2,4,6,10,20,30,40,50 K
\kms\ overlaid on the \hh\ image.  Specific outflows are labeled in Figure
\ref{fig:outflowsh2} on the same scale.} {fig:COonH2}{0.75}

\begin{figure*}[htpb]
    \hspace{-0.6in}
  \includegraphics[scale=0.40,clip=true]{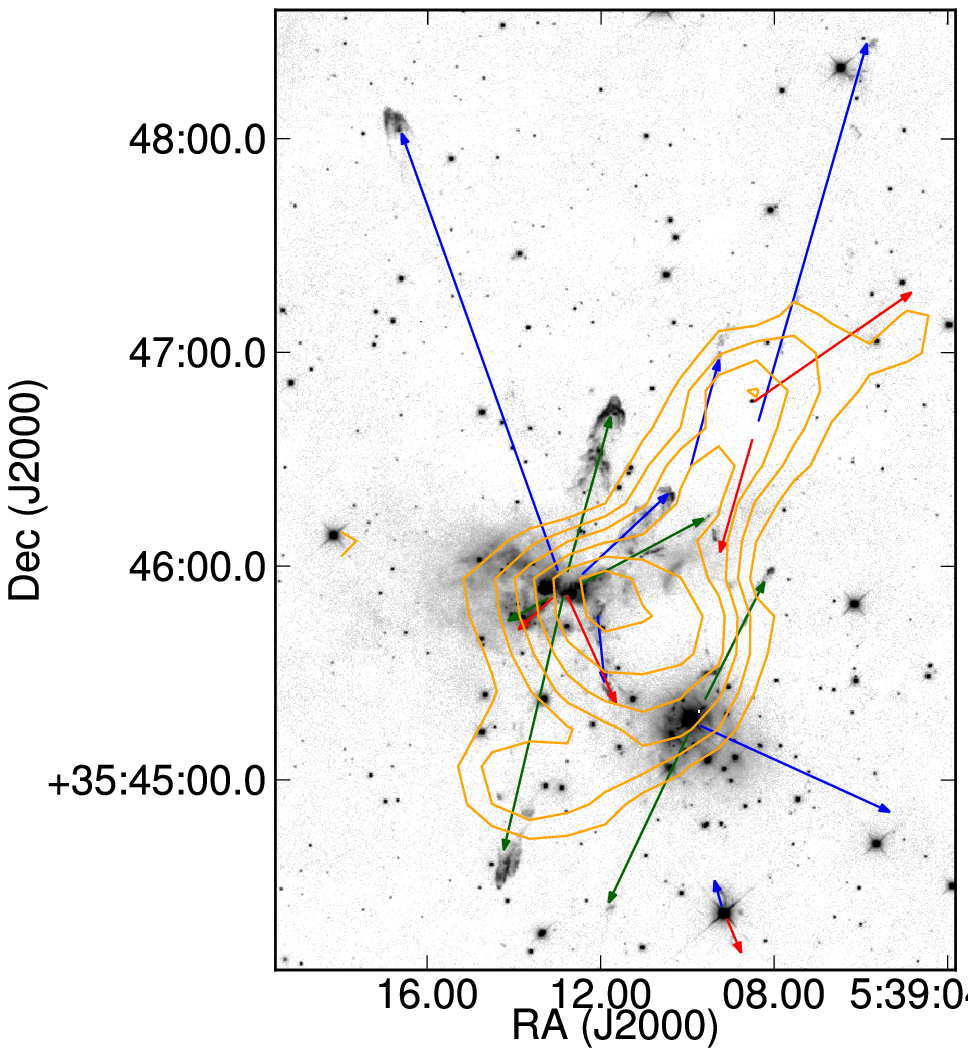}
  \includegraphics[scale=0.40,clip=true]{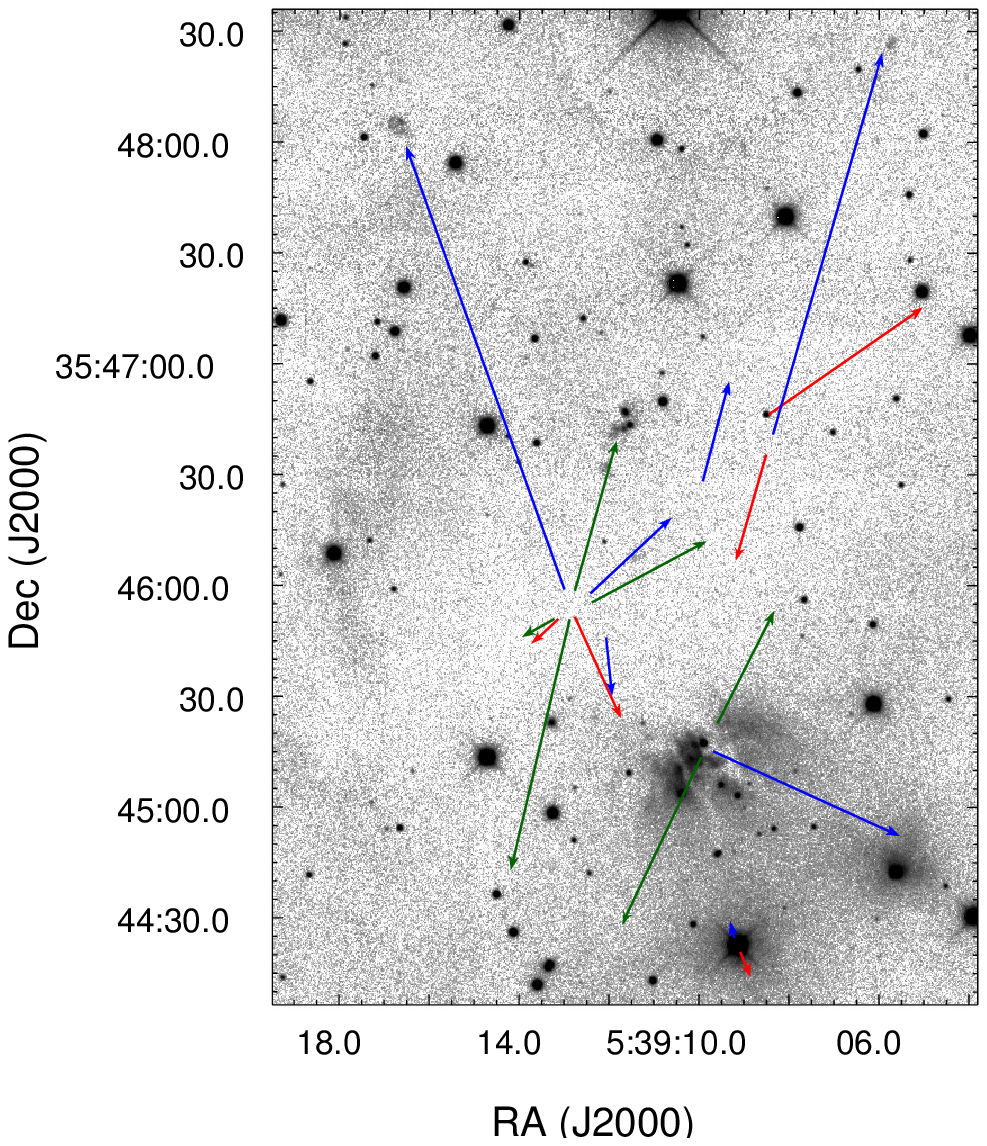}
  \caption{(a) \hh\ image with \so\ peak flux contours at 0.5-1.4 K in intervals of 0.15 K
  overlaid.  With a critical density $\sim3.5\ee{6}$ \citep{leidendb}, this
  transition is a dense gas tracer.  (b) The [S II] image with outflow vectors overlaid.
  Diffuse emission can be seen at the north ends of Outflows 1, 4, and 6 and around the
  reflection nebula near source IR 41.} 
  \label{fig:so_on_h2}
\end{figure*}

\label{sec:outflows}

Figure \ref{fig:outflowsh2} shows the \htwo\ S(1) 1-0 2.1218 \um\ (a rovibrational
transition in the electronic ground state from the $v=1$, $J=3$ to the $v=0$,
$J=1$ state) emission in the vicinity of \region\ with outflows and possible
outflow sources labeled.  The mm cores from \citet{beuther2002} are identified
by red squares.  

The flow vectors in figure \ref{fig:outflowsh2} were chosen on the basis of the
\htwo\ bow shock morphologies and orientations of chains of \htwo\ features,
association with arcsecond-scale CO features on the \citet{beuther2002} Figure
8 CO map, and/or association with lobes of Doppler-shifted CO emission in the
CO 3-2 data.  The color of the vector indicates the suspected Doppler shift;
red and blue correspond to red and blueshifts and green vectors indicate that
the Doppler shift is uncertain.    
 
 {\it \region\ outflow 1:} The most prominent flow in \htwo\ is associated with
 the bright bow-shocks N1 and N6 \citep{khanzadyan2004} located towards PA $\approx$ 345\arcdeg\
 and 170\arcdeg\ respectively from the sub-mm source mms1b \citep{beuther2002}.
 This flow, \citet{beuther2002} outflow A, is associated with redshifted and
 blueshifted CO emission.  The northern shock is seen in \ha\ and [S II]
 emission (figure \ref{fig:so_on_h2}b) and is given a Herbig-Haro designation
 HH 993.

 This flow is indicated by oppositely directed green vectors from the vicinity
 of smm1, 2, and 3.   It is listed as ``Jet 1'' in \citet{qiu2008}.  \citet{kumar2002}
 identified the knot immediately behind the bow shock as a Mach disk.  In the
 \citet{beuther2002} interferometric maps, the north flow contains redshifted features
 and the south flow contains primarily blueshifted features.  There are also blueshifted 
 CO features to the west of the \hh\ knots that are probably part of a different flow
 that is not seen in \hh\ emission.
 
 The velocity of the flow as measured from \hh\ emssion is blueshifted as much
 as 80 \kms (LSR), but one component is blueshifted only 14 \kms\ (see table
 \ref{tab:OutflowH2}), which is consistent with the cloud velocity.  A
 redshifted SiO lobe is present in the south counterflow.  The presence of \ha,
 [S II], and [O III] emission in the north shock and corresponding
 nondetections in the south shock suggest that there is substantially greater
 extinction towards the south knot.  While the velocities in three of the four
 apertures picked along the TripleSpec slit are blueshifted, there are also knots
 with velocities consistent with the cloud velocity.  \citet{porras2000} measure
 the velocity of the counterflow to be -17.3 \kms, which is consistent with 
 the cloud velocity. Outflow 1 is propagating very nearly in
 the plane of the sky.

A line connecting the two bow shocks in Outflow 1 goes directly through
\citet{beuther2007} source mm2a despite the clear association in the
\citet{beuther2002} interferometric CO map (their Figure 8) with mm1a.  The
currently available data do not clarify which is the source of the outflow:
while the bent CO outflow appears to trace Outflow 1 back to mm1a, there are
additional parallel CO outflows towards the confused central region that could
originate from either mm1a or mm2a.

A Spitzer 4.5 \um\ and 24 \um\ source is barely detected in \hh\ 2.5\arcmin\ to
the north of Outflow 1.  It is only apparent when the \hh\ image is smoothed
and would have been dismissed as noise except for the association with
a probably 4.5 \um\ extended source.  It is labeled 24\um\ source 7 in figure
\ref{fig:outflowsh2}.  It appears to be slightly resolved at 4.5\um, and is
therefore likely shocked emission.  The object may be a protostellar source 
with an associated outflow, but its proximity to the projected path of Outflow
1 suggests that it may be an older outflow knot.

{\it \region\ Outflow 2:} The second brightest \htwo\ features trace a bipolar
flow emerging from the immediate vicinity of the sub-mm cluster at PA $\approx$
135\arcdeg\ (red lobe) and 315\arcdeg\ (blue lobe).  It is listed as ``Jet 2'' in
Figure 6 of \citet{qiu2008}.  
The counterflow probably overlaps in the line of sight with the counterflow
from Outflow 3.  It is shorter on the counterflow side either because it has
already penetrated the cloud and is no longer impacting any ambient gas or,
more likely, it has slowly drilled its way out of the molecular cloud and has
not been able to propagate as quickly as the northwest flow.  
The \hh\ velocities measured for these knots are $\sim$ 30 \kms\ blueshifted, or
marginally blue of the cloud LSR velocity.  

The disk identified in \citet{Minier2000} is approximately perpendicular to the
measured angle of Outflow 2 assuming that mm1a is the source of this flow.  It
is therefore an excellent candidate for the outflow source.  A diagram of the
mm1a region is shown in figure \ref{fig:mm1adiagram}.  See Section
\ref{sec:vlaresults} for detailed discussion.

{\it \region\ outflow 3:} The \citet{beuther2002} CO and SiO maps reveal a
third flow, their outflow B at PA $\approx$ 135\arcdeg\  (red lobe) and
315\arcdeg\ (blue lobe).  A chain of  \htwo\ features, \citet{khanzadyan2004}
features N3D and N3E, are probably shocks in this flow.  It is listed as ``Jet
3'' in \citet{qiu2008}.  The two chains of \htwo\ emission indicate that
outflows 2 and 3 are distinct.  There also appears to be a counterflow at a
shorter distance from the mm cores similar to counterflow 2.  

Outflows 2 and 3 may be associated with either redshifted or blueshifted
features in the \citet{beuther2002} CO and SiO maps.  High velocity flows with both
parities are present near both the northwest (\citet{beuther2002} outflow C)
and southeast flow for these jets, but the resolution of the millimeter
observations is inadequate to determine which flow is in which direction.
\citet{porras2000}
measures $v_{LSR} = -7.5$ \kms\ for their knot 4A, which corresponds to the 
blended southeast counterflow of outflows 2 and 3. Their Figure 7 shows a wide 
line that is probably better represented by two or three blended lines, one
consistent with the cloud velocity and the other(s) redshifted.  Since Outflow 2
has a measured blueshift and outflow 3 is significantly fainter, the redshifted
counterflow emission is probably associated with Outflow 2 and the blueshifted
with outflow 3.

{\it \region\ outflow 4:} The JCMT CO data and \htwo\ images reveal a large
outflow lobe consisting of blue lobes 1 and 4 that form a tongue of blueshifted
emission propagating to the northeast at PA $\approx$  20\arcdeg\ (Figure
\ref{fig:outflowsh2}) from the cluster of sub-mm cores.   A faint chain of
\htwo\ features runs along the axis of the CO tongue and terminates in a bright
\htwo\ bow shock located at the northern edge of \ref{fig:outflowsh2}.   Several
\htwo\ knots lie along the expected counterflow direction, but that portion of
the field contains multiple outflows and is highly confused.  If the
counterflow is symmetric with the northeast knot, it extends 2.1 parsecs on the
sky.  

The bow shock of Outflow 4 is seen in the HII and [S II] images, implying that
the extinction is much lower than in the cluster.  Two apertures placed along
the bow shock reveal that it is blueshifted about 70\kms\ and may be extincted
by as little as $A_V\sim.5$.  It is designated HH 994.
 
{\it \region\ outflow 5:} Figure \ref{fig:outflowsh2} shows a bright chain of
\htwo\ knots and bow shocks starting about 10\arcsec\ west of mm3 and
propagating south at PA $\approx$ 190\arcdeg.  The SiO maps of
\citet{beuther2002} show a tongue of blueshifted emission along this chain
(their Outflow C).   The outflow projects back to H$^{13}$CO$^+$ source 3,
which is also a weak mm source.  A lack of obvious counterflow and the
possibility that the knots identified with Outflow 5 could be associated with a
number of different crossing flows makes this identification very tentative.
Higher spatial resolution observations will be required to determine the
association of this outflow.

 {\it \region\ outflow 6:} The fourth brightest source in the Spitzer 24\um\ 
 data is located at J(2000) = 05:39:08.5, +35:46:38 (source 5 in the \region\ 
 section of the \citet{qiu2008} catalog, referred to in table \ref{tab:OutflowH2} 
 as Q5) in the middle of the
 molecular ridge that extends from \region\ towards the northwest (24\um\ 
 object 4 in figure \ref{fig:outflowsh2}).   The star is
 located at the northwest end of the tongue of 1.2mm emission mapped by
 \citep{beuther2002} with the MAMBO instrument on the IRAM telescope.  This
 part of the cloud is also seen in silhouette against brighter surrounding
 emission at 8\um.  At wavelengths below 2\um, it is fainter than 14-th
 magnitude and therefore is not listed in the 2MASS catalog, and it is not
 detected in \citet{yan2009} down to 19th magnitude in K.
 
 Spitzer data indicates very red colors between 3.6 and 70 \um, indicating that
 this object is likely to be a Class I protostar.  The SED is fit using the
 online tool provided by \citet{robitaille2007}.  Unfortunately, a wide variety
 of parameters all achieved equally good fits, so no conclusions are drawn
 about the stellar mass or other very uncertain parameters.  However, the top
 models all had $A_V > 20$ and many in the range 30-50, indicating that the
 line of sight is probably through a thick envelope or disk towards this
 source.  

 
 This source lies at the base of the tongue of blueshifted CO 3-2
 emission that extends northwest of \region\ at PA $\approx$ 345\arcdeg\  and
 has mass $\sim .5\msun$.  A pair of \htwo\ features,
 \citet{khanzadyan2004} N12A and N12B are located 30 and 55\arcsec\ from the
 suspected YSO, forming a chain along the axis of the blueshifted CO
 tongue.    \citet{khanzadyan2004} \htwo\ knot N3F lies along the flow axis in
 the redshifted direction.
 
{\it \region\ outflow 7:} The 20\arcsec\ long chain of \htwo\ knots labeled
\citet{khanzadyan2004} N11 appears to trace part of a jet at PA $\approx$
345\arcdeg\ that propagated parallel to outflow 6 about 20\arcsec\ to the east.
The northwest portion of Outflow C in the \citet{beuther2002} SiO map is in 
approximately the same direction as Outflow 7, and it may represent a redshifted
counterflow to the northwest-pointing \hh\ knots.
The jet axis passes within a few arc-seconds of a faint and red YSO located at
J(2000) = 05 39 10.0, +35 46 27 (blue diamond in figure \ref{fig:outflowsh2}
about 35\arcsec\ south of the southern end of the \htwo\ feature).  It may be a
24\um\ source but is lost in the PSF of the bright source at the center of
\necluster.  This object is also undetected down to 19th magnitude in the
\citet{yan2009} K-band image.
 
{\it \region\ outflow 8:}   A prominent jet-like \htwo\ feature protrudes from
the vicinity of \swcluster\ at PA $\approx$ 335\arcdeg\  and ends in
 bright knot 
N9.     The feature N5B is is
located just outside the ring of \htwo\ emission that surrounds the IRAS source
at the base of the jet.  Towards the southeast, 
knot N6 is located opposite knot N9 with respect to the southwest cluster.  IR 41,
the \ha\ emission source, labeled 24\um\ source 6 in figure \ref{fig:outflowsh2},
is probably the source of this outflow.
 
{\it \region\ outflow 9:} In the Spitzer and K$_s$ images, an infrared
reflection nebula opens towards the southwest at PA $\approx$ 245\arcdeg\ and
points towards a blueshifted CO region.  The reflection nebula is also seen in
\ha.  It is likely that the CO emission in CO Region 1 (table \ref{tab:comeas})
traces a fossil cavity whose walls provide the scattering surface of the
reflection nebula.

{\it \region\ outflow 10 and IR 6:}   A bright \htwo\ filament protrudes at PA $\approx$
15\arcdeg\ towards the northeast of IR 6 (24\um\ source 1, \citet{qiu2008}
source 8).  The star is the third brightest 24\um\ source in the \region\
region.   Since it is visible at visual wavelengths, it is not heavily
embedded.     Its  H$\alpha$ emission and association with an outflow lobe and
\htwo\ emission suggest that it is a moderate mass Herbig AeBe star associated
with the \region\ complex.  The optical spectrum confirms this hypothesis: the
star has \ha\ absorption wings on either side of a very bright, asymmetric \ha\
emission profile (see section \ref{sec:dis}). 

IR 6 is seen to be the source of Outflow 10.  Data for this source is available
from $\sim$0.45-24\um, so the \citet{robitaille2007} spectral fitter puts
strong constraints on the star's mass and luminosity.  The measured mass and
luminosity are  $M=4.5\pm0.5$ \msun\ and $L = 10^{2.3\pm.25} L_\odot$, parameters
consistent with a B7V ($\pm 1$ spectral class) main sequence star.  The range
of ages in the models covers $10^4-10^7$ years but favors stars in the range
$10^5-10^6$ years.

While there is a small clump of redshifted CO emission to the northeast of the
object, the \htwo\ spectrum shows that the north flow is blueshifted $v_{LSR}\sim-40
$\kms, and the lack of a visible counterflow suggests that the counterflow may
be masked behind an additional extincting medium.  The counterflow drawn in
figure \ref{fig:outflowsh2} is not seen in emission but is identified as a
probable location for a counterflow because of the confident association of
outflow 10n with source IR 6.

{\it \region\ outflow 11:} A chain of \hh\ knots is seen at 2.12\um\ and in the
Spitzer 4.5\um\ image.  They trace back to either IR 78 or 24\um\ source 4.
There is a tongue of redshifted CO 3-2 emission in the same direction as this
flow that suggests it may be redshifted.

{\it IR 41}: 
There is an arc-like \hh\ emission feature surrounding the \ha\ emission line
star IR 41.  This implies that the star is probably a late B-type star with too
little Lyman continuum emission to generate a photon-dominated region (PDR) but
enough soft UV to excite \hh.  From the measured \ha\ and nondetection of \hb\
at the star's location down to a 5-$\sigma$ limit of 1\ee{-17} erg s$^{-1}$
\persc \AA$^{-1}$, a lower limit on the extinction column $A_V=15$ is derived.
The \citet{robitaille2007} fitter yields a mass estimate of 7.4$\pm 0.6$\msun
and luminosity $L=10^{2.97\pm.16}L_\odot$ among the 222 best fits out of a grid
of 200,000 model SEDs (fits with $\chi^2<5000$).  The luminosity is very well
constrained, varying only modestly to $L=10^{2.99\pm.15}L_\odot$ for the 904
best fits ($\chi^2 < 10000$),  while the mass shifts down to $6.5\pm1.0\msun$.
The mass estimate may be biased by the lower number of high-mass models
computed.  The star's mass is most compatible with a main sequence B4V star,
though its luminosity is closer to a B5V star.  The disk mass is constrained to
be $>10^3 \msun$.  The age is reasonably well constrained to be $T =
10^{5.78\pm.12}$ for the best 904 models, but is essentially unconstrained for
the best 222.  Similarly, the stellar temperature is entirely unconstrained by
the fitting process.

The very high values of $\chi^2$ would normally be worrisome, but the $\chi^2$
statistic only represents statistical error, while the data is dominated by
various systematic errors including calibration offsets in the optical/NIR and
poor resolution in the far-IR.  Therefore, it is not possible to find a perfect
model fit, but still possible to put constraints on the physical properties of
the source.

\Figure{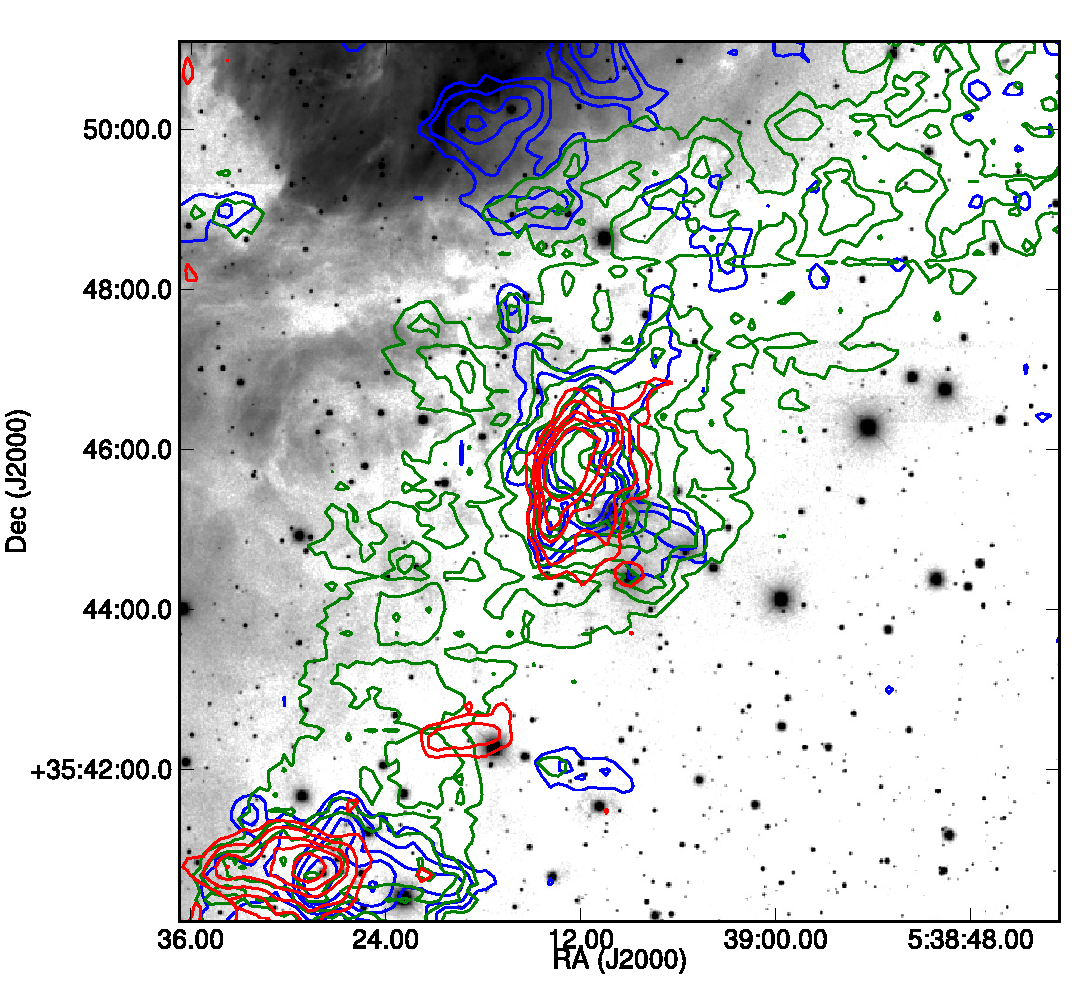}{The \ha\ image with CO contours at
redshifted, blueshifted, and middle velocities in red, blue, and green
respectively.  Contours are at 2,4,8,12,20 K \kms\ for the red and blue, 
and 20,25,30,40,50,60,70 K \kms\ for the green.  Red is integrated from
-12 to -4 \kms, Blue from -31 to -21 \kms, and green from -21 to -12 \kms.
}{fig:HA_with_CO}{1.0}

{\it South of \region}: 
There is a symmetric flow with one faint \hh\ knot and a bright central source
about 4\arcmin\ south of \region.  The \hh\ knot is at J(2000) = 05:39:15.63 +35:42:13.2.
The flow has a clear red and blue region as identified in figure
\ref{fig:cofig}; the red flow extends from -9 to -14 \kms\ and the blue from -19
to -23 \kms\ (the outflow is swamped by ambient emission in the intermediate
velocity range).  The outflow is $\sim 2\arcmin$ long, though the probable
source identified is not directly between the two lobes.  The ellipses used are
labeled in table \ref{tab:comeas} as Red S and Blue S.

\subsection{Imaging results: Optical}
Deep [S II] images show that some of the outflows have pierced through the
obscuring dust layers or excited extremely bright sulfur emission.
\citet{khanzadyan2004} knot N1 at the end of Outflow 1 is visible [S II] emission
The bow of outflow 4 and the northwest end of outflow 6 are
detected in [S II].  Only the Outflow 1 and 4 bows are detected in \ha\ 
emission, indicating that the emission is most likely from shock heating, not
external photoionizing radiation.  If the shocks were externally irradiated, we
would expect the emission to be dominated by the recombination lines.  Because
they have been detected in the optical, these two flows can be classified
as Herbig-Haro objects.

\subsection{CO results}

\region\ is located at the center of the CO 3-2 integrated velocity maps
(Figure \ref{fig:cofig}).  The parent molecular cloud, centered at $v_{LSR} =
-17.5$ \kms ,  extends from the southeast towards the northwest with the
brightest emission coming from the core associated with \swcluster, while the
highest integrated emission is associated with \necluster.  \necluster\ has a central
velocity of $\sim -16.0$ \kms\ from the optically thin \ceighteeno\ 2-1 measurements.
Material that has been swept up and accelerated by jets and outflows can be
seen at velocities $v_{LSR} < -21$ \kms and $v_{LSR} > -12$ \kms\ (Figure
\ref{fig:cofig}).  The integrated CO 3-2 map peaks at J(2000) = 05:39:12.8
+35:45:55, while the highest observed brightness temperature is at J(2000) =
5:39:09.4 +35:45:12.  This offset is discussed in the context of CO isotopologues in
section \ref{sec:co21} and in section \ref{sec:discussion-outflows}.

Regions with line wings relative to the ambient cloud within 5\arcmin\ of the
northeast cluster were assumed to be associated with outflows from the cluster.
Further than 5\arcmin, it is likely that the high velocity wings are accelerated by
neighboring HII regions (see section \ref{sec:surroundings}).  These line
wings were integrated over the velocity range -34 to -21 \kms\ (blue) and -12 to
1 \kms\ (red) to acquire estimates of the outflowing mass under the assumption
that outflowing gas is optically thin.  The extracted regions are displayed in
Figure \ref{fig:cofig}b and measurements in table \ref{tab:comeas}.  The line
wings in the central arcminute and central 5 arcminutes were measured for
comparison with lower resolution data and to compute a total outflow mass in
the central region.  

The objects in Table \ref{tab:comeas} labeled CO Region 1, 2, and 3 have
uncertain associations with outflows.  CO Region 1 is tentatively associated
with outflow 11. CO region 2 may be associated with Outflow 3 but is in a
highly confused region and may have many contributors.  CO region 3 is probably
associated with outflow 10.  In contrast, the associations with outflows 4 and
6/7 are more certain because they are further from the central region and less
confused.   Outflow 1 is seen at high velocities in \citet{beuther2002}
interferometer maps.  Outflow 9 is selected primarily based on CO emission.

\Figure{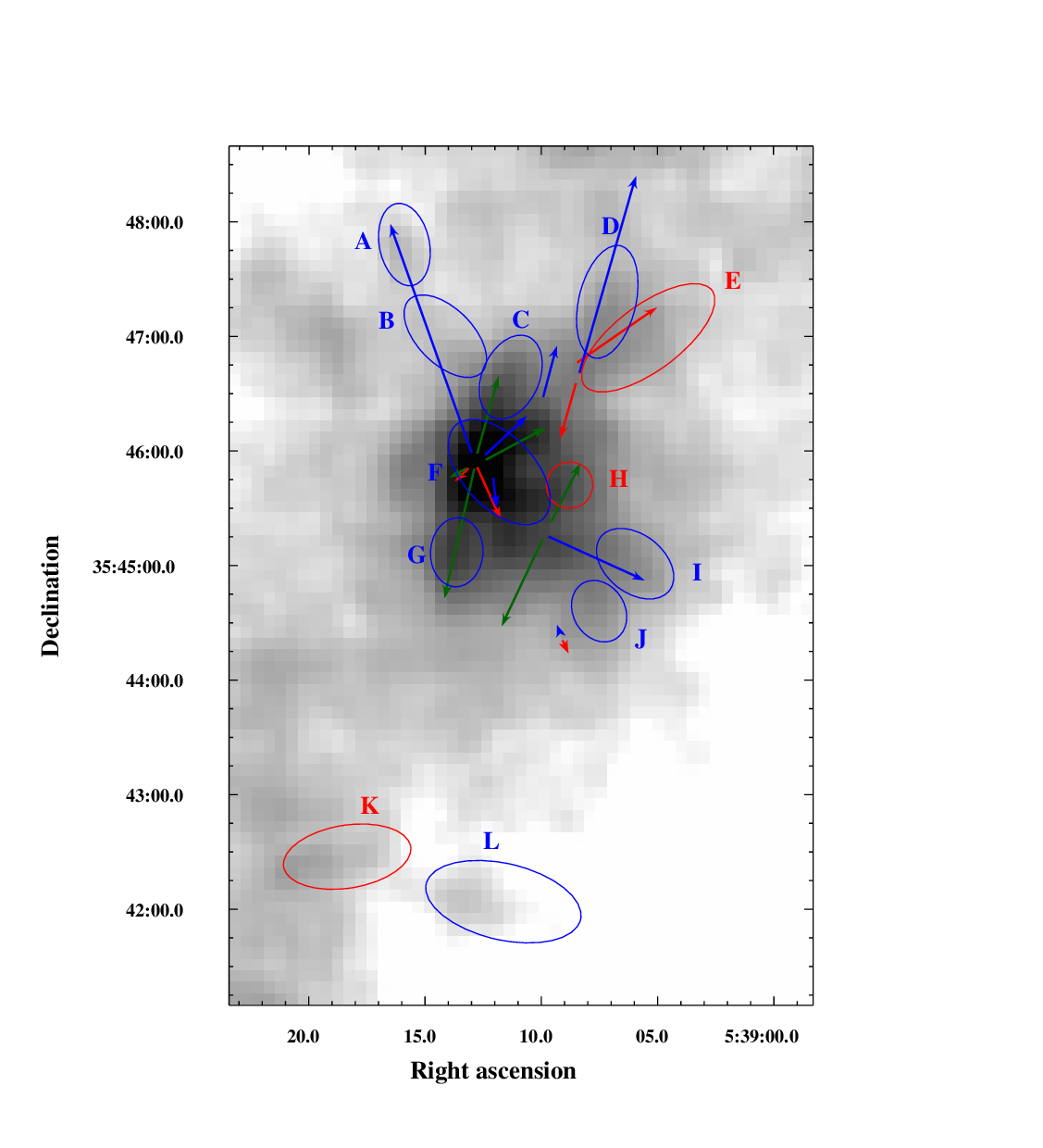}
{
The JCMT HARP CO J=3-2 map integrated over all velocities with significant
emission (-34 \kms\ to -4 \kms) shown in gray log scale from 0 to 150 K \kms.
The elliptical regions over which line wings were integrated are shown with
blue and red circles corresponding to blue and red line wings.  The
measurements are presented in table \ref{tab:comeas}.}{fig:cofig}{1.0}

\Figure{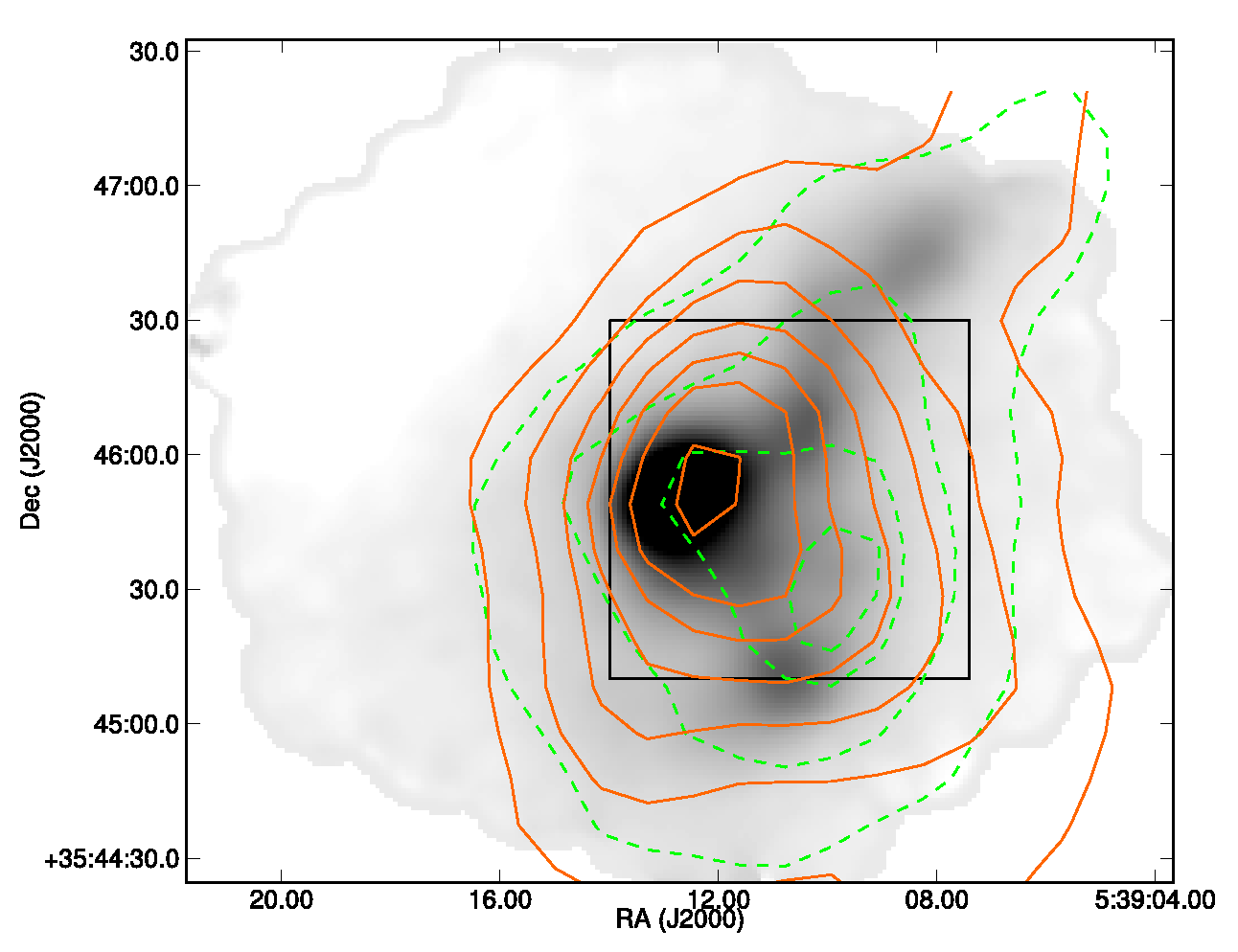}{SCUBA 850\um\ image in linear grayscale from
-1 to +10 mJy/beam, with a saturated peak of 24 mJy/beam, with \twelveco\ 2-1
(orange solid, contours at 45,60,85,100,115,130,145 K \kms) and \thirteenco\
2-1 (green dashed, contours at 20,30,40,45 K \kms) integrated contours.  The
box shows the region plotted in Figure \ref{fig:co21map}.}{fig:scuba_co21}{1.0}
\Figure{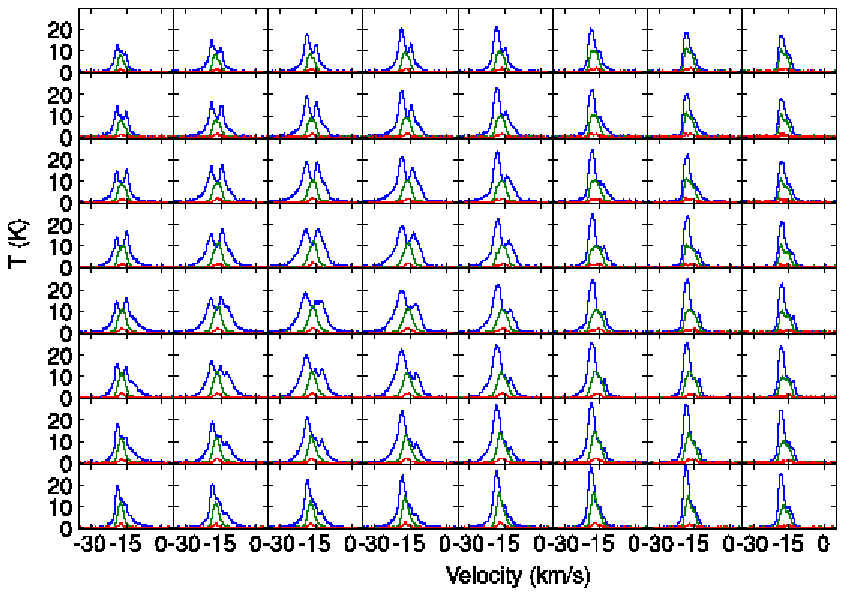}{CO spectra of \necluster\ in \twelveco\ (blue),
\thirteenco\ (green), and \ceighteeno\ (red).  The top-left plot is the pixel
centered at J(2000) = 5:39:13.67 +35:46:26.0 and each pixel is 10\arcsec\ on a side. 
The region mapped here is shown with a box in Figure \ref{fig:scuba_co21}.
Redshifted self-absorption, a possible infall tracer, is evident in the \twelveco\
spectra in the outer pixels.  The inner pixels show self-absorption only at
central velocities: this may be an indication that emission from outflows
dominates any infall signature, or simply that there is no bulk infall towards
\necluster.}{fig:co21map}{0.75} \Figure{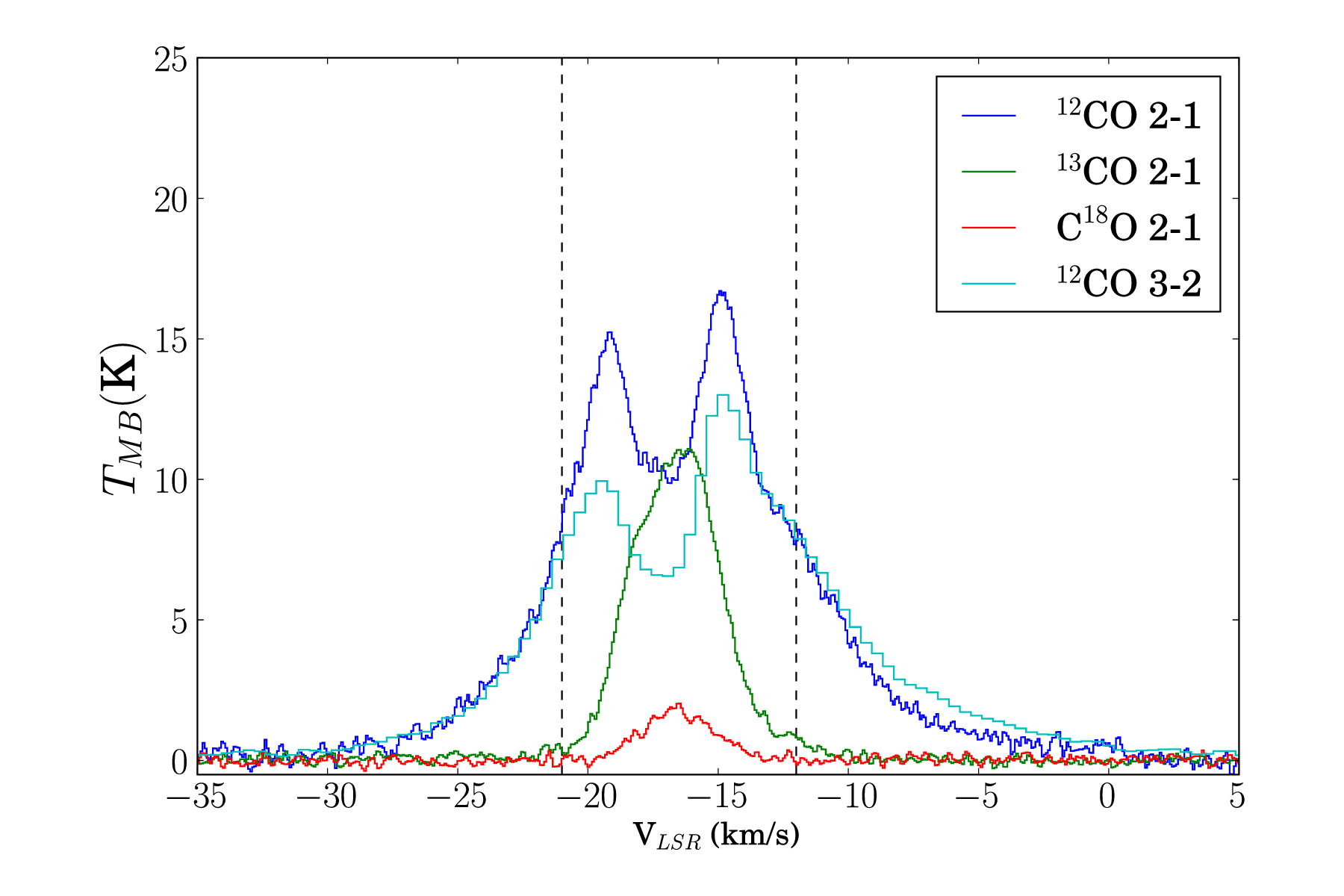}{CO spectra of inner
12\arcsec\ centered on \necluster\ for all observed CO lines.   The CO 3-2 and
2-1 beams are not matched, but in both cases the area integrated over is 1-2
resolution elements across.  The divisions demarcating the red and blue line
wings are shown with vertical dashed lines at $v_{LSR}=-21$ and -12 \kms.
}{fig:co21_all3}{1.0}

\begin{deluxetable}{lccccc}
  \tabletypesize{\footnotesize}
    \centering
    \tablecaption{Measured properties of CO flows
      \label{tab:comeas}
    }
\tablehead{
\colhead{\tablenotemark{a}Region Name} & \colhead{$\int T_{mb}*$} & \colhead{$M (M_\odot)$} & \colhead{p ($M_\odot$ \kms)} & 
\colhead{N (\persc)} & \colhead{E ($10^{42}$ erg)}}
    \startdata
\tablenotemark{b}A. Outflow 4a    &      4.27  &     .022  &     .15  &  1.4\ee{19}  & 11 \\
\tablenotemark{b}B. Outflow 4b   &      4.60 & .032 & .21 &   1.5\ee{19}  & 13 \\
\tablenotemark{b}C. Outflow 1n   &      14.5 & .088 & .71 & 4.8\ee{19} & 66 \\ 
\tablenotemark{b}D. Outflow 6/7  &       4.45 & .045 & .30 & 1.5\ee{19} & 29 \\
\tablenotemark{r}E. CO Region 3      &    1.31 & .016 & .112 & 4.3\ee{18} & 8.5 \\
\tablenotemark{b}F. \necluster\ &      41.8 & .464 & 3.72 & 1.4\ee{20} & 330 \\
\tablenotemark{m}F. \necluster\ &      132.9& 1.47 & -    & 4.4\ee{20} & -\\
\tablenotemark{r}F. \necluster\ &      30.0 & .333 & 2.03 & 9.9\ee{19} & 135 \\
\tablenotemark{b}G. Outflow1s   &      14.6 & .064 & .48 & 4.8\ee{19} & 40 \\
\tablenotemark{r}H. CO Region 2      &    4.54 & .012 & .074 & 1.5\ee{19} & 5 \\
\tablenotemark{b}I. Outflow 9    &      6.33 & .039 & .39 & 2.1\ee{19} & 43\\
\tablenotemark{b}J. CO Region 1  &     3.61 & .015 & .12 & 1.2\ee{19} & 11\\
\tablenotemark{r}K. Red S   &     5.26 & .051 & .34  & 1.7\ee{19} & 26 \\
\tablenotemark{b}L. Blue S  &     3.66 & .053 & .47  & 1.2\ee{19} & 47 \\
\tablenotemark{b}1\arcmin\ aperture\tablenotemark{c}& 15.1 & .96 & 7.6 & 5.0\ee{19} & 670\\
\tablenotemark{b}3\arcmin\ aperture& 2.7  & 1.6 & 12 & 9.0\ee{18} & 1000\\
\tablenotemark{b}5\arcmin\ aperture& 1.7  & 2.7 & 20 & 5.6\ee{18} & 1600 \\
\tablenotemark{r}1\arcmin\ aperture & 11.8 & 0.75 & 4.7 & 3.9\ee{19} & 320\\
\tablenotemark{r}3\arcmin\ aperture & 1.9 & 1.1 & 6.8 & 6.2\ee{18} & 460\\
\tablenotemark{r}5\arcmin\ aperture & 0.96 & 1.5 & 10 & 3.2\ee{18} & 640\\
\tablenotemark{b}1\arcmin\ $^{12}$CO 2-1 & 10.4 & .94 & 7.1 & 4.9\ee{19} & 590 \\
\tablenotemark{m}1\arcmin\ $^{12}$CO 2-1 & 97.78 & 8.83 & - & 4.6\ee{20} & - \\
\tablenotemark{r}1\arcmin\ $^{12}$CO 2-1 & 9.17 & 0.83 & 5.52 & 4.3\ee{19} & 430 \\
\tablenotemark{m}1\arcmin\ $^{13}$CO 2-1 & 41.12 & 211  & - & 1.1\ee{22} & - \\
\tablenotemark{m}1\arcmin\ C$^{18}$O 2-1 & 5.31 & 271 & - & 1.4\ee{22} & - \\
\enddata
\tablenotetext{a}{Unless labeled otherwise, regions are extracted from CO 3-2 data as shown in figure \ref{fig:cofig}b }
\tablenotetext{b}{Blue integration over velocity range -34 to -21 \kms}
\tablenotetext{c}{Apertures are centered on J(2000) = 05:39:11.238 +35:45:41.80 in \necluster}
\tablenotetext{r}{Red integration over velocity range -13 to -4 \kms}
\tablenotetext{m}{Middle range integration over -21 \kms\ to -13 \kms.  Assumed not to be outflowing, so no momentum is computed}
\end{deluxetable}

\subsection{Near-infrared spectroscopy: Velocities}
\label{sec:tspecresults}

The slit positions used and apertures extracted from those slits are displayed
in Figure \ref{fig:tspecslits}.  Position-velocity diagrams of the 1-0 S(1)
line are displayed in Figure \ref{fig:outflows_h2_pv}.  Velocity measurements
are presented in Table \ref{tab:OutflowH2}.

\Figure{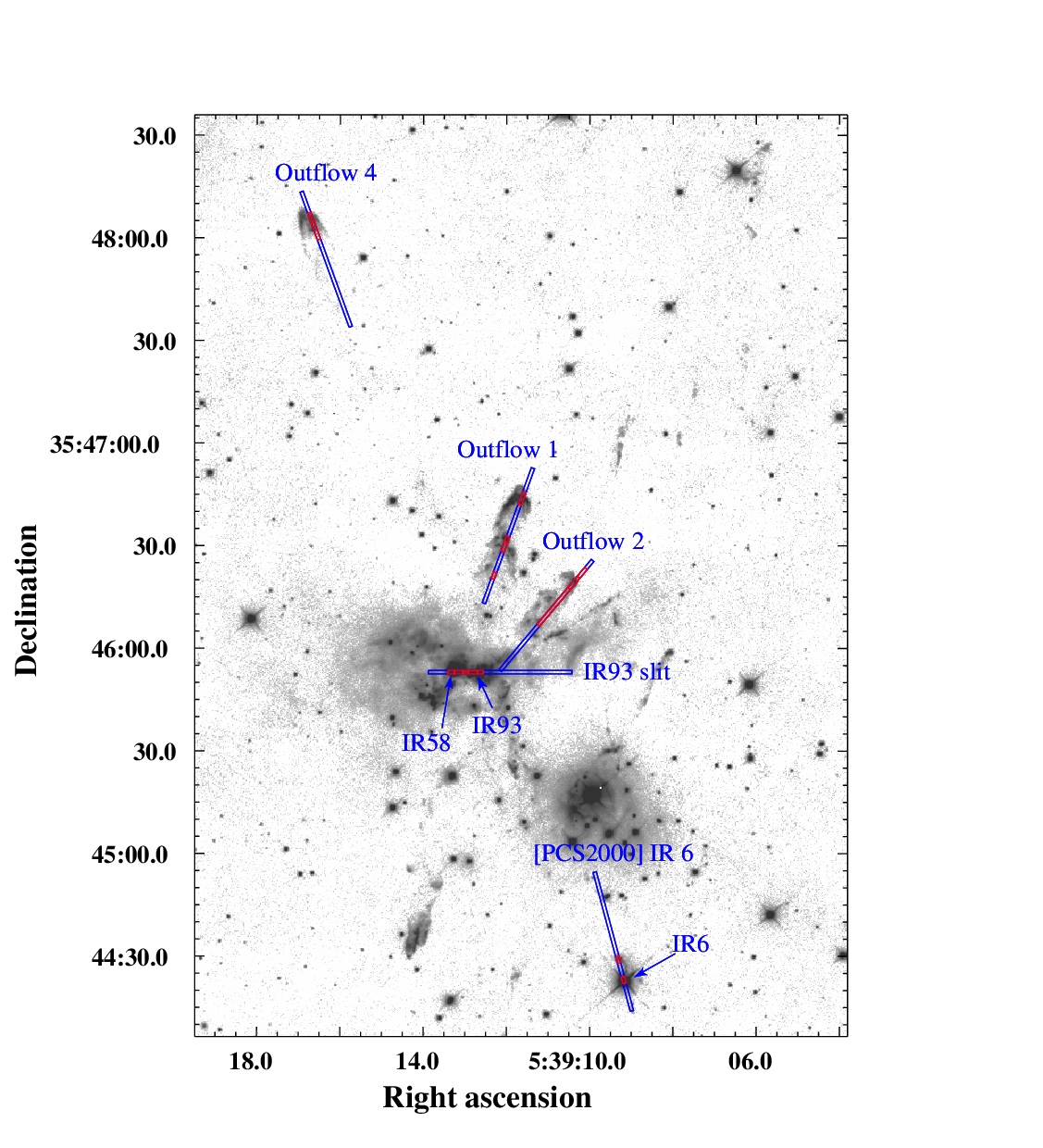}{TripleSpec slits (blue) overlaid on the
\hh\ image.  The red boxes indicate the apertures extracted from those slits to
fit and measure \hh\ properties.  The apertures are also indicated in the
position-velocity diagrams.}{fig:tspecslits}{0.75}

The near-IR spectrum of Outflow 1 has the largest signal.  All of the K-band
\hh\ lines except the 2-1 S(0) 2.3556 \um\ (too weak) and 1-0 S(4) 1.8920 \um
(poor atmospheric transmission) lines were detected (see Table
\ref{tab:nirmeas}).  Velocities from gaussian fits to each line are reported.
In the central portion of \necluster, outflowing \hh\ emission at
$v_{LSR}\approx-30$ \kms\ is detected.  This material may be associated  with a
line-of-sight flow, or may originate from the base of the already identified
flows 1-3.  In source IR 58,  Br$\gamma$ and He I 2.05835\um are detected,
indicating that there is an embedded PDR in this source.  There is a hint of a
second, fainter star adjacent to IR 58.  IR 93 is observed to be a double
source in the TripleSpec spectrum, but the spectrum is too weak to do any
identification.  Br$\gamma$ and possibly He I are detected at fainter levels.

Table \ref{tab:nirmeas} shows the measured line strengths (when detected) of
all \hh\ lines in each aperture.  The errors listed are statistical errors
that do not include the systematics errors introduced by a failure to correct
for narrow atmospheric absorption lines.

\Figure{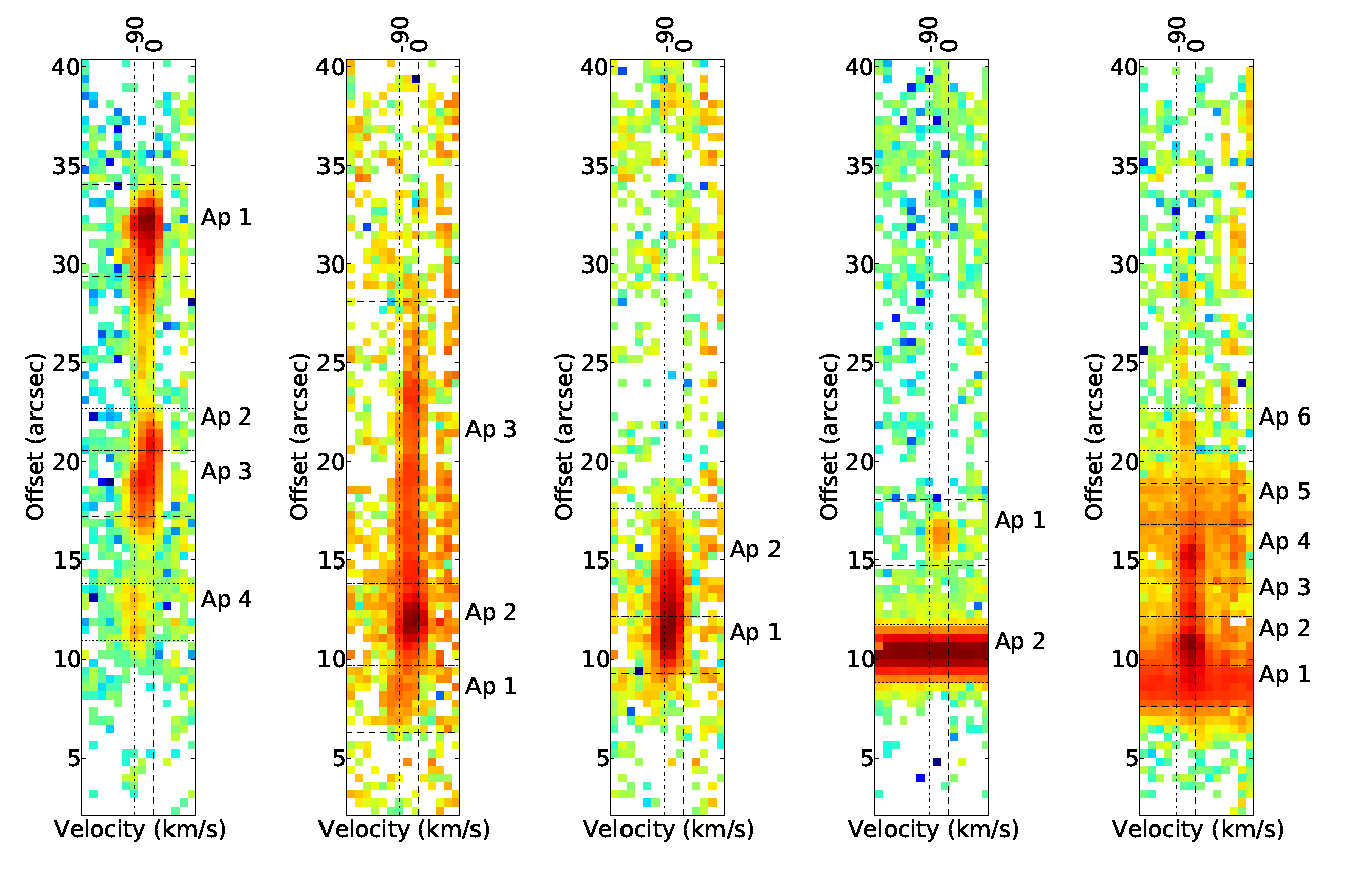}{Position-velocity diagrams of the \hh\ 2.1218 \um\ line in Outflows
1,2, 4, IR 6, and IR93/IR58.  The velocity range is from -340 to 190 \kms.}{fig:outflows_h2_pv}{1.0}
  
\Table{cccccc}
{TripleSpec fitted \htwo\ outflow velocities}
{Outflow Number & Aperture Number &  \tablenotemark{a}v$_{LSR}$ (\kms) &\tablenotemark{b}v$_{LSR} (\kms)$ }
{tab:OutflowH2}
{
1    & 1 & -33.54 (0.15)    & -31.85 (0.32) \\ 
1    & 2 & -13.60 (0.57)    & -13.56 (0.96) \\ 
1    & 3 & -40.51 (0.41)    & -36.13 (0.81) \\ 
1    & 4 & -88.7  (2.8)     & -83.7  (7.9)  \\ 
2    & 1 & -82.6  (7.6)     & -81    (21)   \\ 
2    & 2 & -30.41 (0.57)    & -28.9  (1.4)  \\ 
2    & 3 & -33.89 (0.62)    & -35.2  (3.7)  \\ 
4    & 1 & -73.34 (0.48)    & -70.2  (1.1)  \\ 
4    & 2 & -64.08 (0.61)    & -67.8  (2.2)  \\ 
IR6  & 1 & -39.4  (1.6)     & -39.4  (4.2)  \\ 
IR93 & 2 & -26.07 (0.43)    & -26.85 (0.97) \\ 
IR93 & 3 & -30.6  (1.5)     & -32.0  (2.5)  \\ 
IR93 & 4 & -29.14 (0.77)    & -30.3  (2.1)  \\ 
IR93 & 6 & -47.7  (7.9)     & -71    (37)   \\ 
}{
\tablenotetext{a}{Measured from \hh\ 1-0 S(1) 2.1218 \um\ line}
\tablenotetext{b}{Measured from all detected \hh\ lines fit with model described in section \ref{sec:tspecresults}}
}

\clearpage
\begin{deluxetable}{ccccccccc}
    \centering
  \tabletypesize{\scriptsize}
    \tablecaption{Measured properties of \hh\ flows}
    \tablehead{
    Outflow  & \tablenotemark{a}Center  & \tablenotemark{b}PA  & \tablenotemark{c}Length  & 
    \tablenotemark{d}Source  & \tablenotemark{e}Flow  & \tablenotemark{e}Counterflow  & \tablenotemark{f}Age  & \tablenotemark{g}LOS  \\
    & & & & & Length & Length & (50 \kms) & Velocity \\
    }
\startdata
1 &        05:39:13.023 +35:45:38.66 & -13.3     &142.3"   &mm2?     &58         &84.2        &1.4e4         &-    \\
2 &        05:39:13.058 +35:45:51.28 & -47.0     &44.6"    &mm1a     &44.6       &-           &6.6e3         &Blue \\
3 &        05:39:12.48  +35:45:54.9  & -62       &44"      &mm3?     &44         &-           &6.5e3         &Red  \\
4 &        ambiguous                 & 17.8-21.8 &141-144" &  ?      &141-144    &-           &2.1e4         &Blue \\
5 &        05:39:12     +35:45:51    & 170       &38-48    &mm3?     &38-48      &-           &6.5e3         &Blue \\
6 &        05:39:09.7   +35:45:17    & 14.5      &197      &Q5       &197        &-           &2.9e4         &Blue \\
8 &        05:39:10.002 +35:45:10.87 & -154.6    &105.5"   &IR41?    &54.7       &52.9        &7.9e3         &-    \\         
\enddata
\tablenotetext{a}{Midpoint of bipolar outflow if symmetric, position of jet source candidate if asymmetric}
\tablenotetext{b}{Position angle uncertainties are $\sim 5\degree$ because they are not perfectly collimated,
causing an ambiguity in their true directions.  The exact angles used to draw vectors in figure \ref{fig:outflowsh2}
are listed for reproducibility.}
\tablenotetext{c}{Total length of outflow on the sky, including counterflow}
\tablenotetext{d}{Candidate jet source object.  Outflows 2 and 6 have clear associations,
the others are weaker candidates.}
\tablenotetext{e}{Flow length is the distance from the CENTER position to the last \hh\ knot 
in the position angle direction as listed.  Counterflow length is the distance from the CENTER position
to the opposite far knot.}
\tablenotetext{f}{Timescale of jet assuming it is propagating at 50 \kms, an effective lower limit to see
\hh\ emission.  If two lengths are available, uses the longer of the two.  These are lower limits to the 
true timescale \citep{parker1991}.}
\tablenotetext{g}{The parity of the outflow along the line of sight.  Outflow 1 and 8 have counterflows with
parities as indicated in figure \ref{fig:outflowsh2}}
\end{deluxetable}

\clearpage
\begin{deluxetable}{ccccccccccccc}\setlength\tabcolsep{3pt}
  \scriptsize
  \tabletypesize{\tiny}
  \tablewidth{0pt}
    \centering
    \tablecaption{Measured \hh\ line strengths
      \label{tab:nirmeas}}
    \tablehead{
        &1-0 S(0)&1-0 S(1)&1-0 S(2)&1-0 S(3)&1-0 S(6)&1-0 S(7)&1-0 S(8)&1-0 S(9)&1-0 Q(1)&1-0 Q(2)&1-0 Q(3)&1-0 Q(4)  \\
aperture&2.2233&2.12183&2.03376&1.95756&1.78795&1.74803&1.71466&1.68772&2.40659&2.41344&2.42373&2.43749}
\startdata
outflow1ap1  &   3.60E-15&  9.80E-15&  5.50E-15&  1.20E-14&  4.70E-15&  3.10E-15&  8.60E-16&  1.10E-15&  9.20E-15&  6.10E-15&  1.10E-14&  6.90E-15 \\
             & ( 2.4e-17)&( 3.4e-17)&( 6.8e-17)&(   2e-16)&(   2e-16)&( 2.8e-17)&( 2.8e-17)&( 2.7e-17)&( 1.4e-16)&( 7.4e-17)&( 8.8e-17)&( 7.8e-17) \\
outflow1ap2  &   7.10E-16&  1.80E-15&  9.90E-16&  1.80E-15&         -&         -&         -&         -&  3.00E-15&  1.90E-15&  3.20E-15&  2.00E-15 \\
             & ( 2.1e-17)&( 2.7e-17)&( 6.8e-17)&( 1.7e-16)&         -&         -&         -&         -&( 1.3e-16)&(   4e-17)&( 7.8e-17)&( 3.8e-17) \\
outflow1ap3  &   1.60E-15&  4.10E-15&  2.20E-15&  4.70E-15&         -&  8.30E-16&         -&         -&  5.60E-15&  3.70E-15&  6.60E-15&  4.80E-15 \\
             & ( 2.4e-17)&( 3.4e-17)&( 6.3e-17)&( 1.8e-16)&         -&( 2.8e-17)&         -&         -&( 1.4e-16)&( 5.9e-17)&( 8.2e-17)&( 5.2e-17) \\
outflow1ap4  &          -&  9.00E-16&         -&         -&         -&         -&         -&         -&         -&         -&         -&         - \\
             &          -&(   3e-17)&         -&         -&         -&         -&         -&         -&         -&         -&         -&         - \\
outflow2ap1  &          -&  3.60E-16&         -&         -&         -&         -&         -&         -&         -&         -&         -&         - \\
             &          -&( 1.5e-17)&         -&         -&         -&         -&         -&         -&         -&         -&         -&         - \\
outflow2ap2  &   9.40E-16&  2.40E-15&  1.50E-15&  1.80E-15&         -&  4.00E-16&         -&         -&  3.00E-15&         -&  3.70E-15&         - \\
             & ( 1.7e-17)&( 2.2e-17)&( 5.8e-17)&( 1.1e-16)&         -&( 2.3e-17)&         -&         -&( 4.7e-17)&         -&( 7.9e-17)&         - \\
outflow2ap3  &   2.10E-15&  1.90E-15&  1.80E-15&  2.20E-15&         -&  6.70E-16&         -&         -&  5.70E-15&         -&  7.30E-15&         - \\
             & ( 1.7e-17)&( 2.2e-17)&( 5.1e-17)&( 1.3e-16)&         -&( 2.9e-17)&         -&         -& -8.00E-16&         -& -8.00E-16&         - \\
outflow4ap1  &   5.50E-16&  2.00E-15&  8.50E-16&  2.00E-15&         -&  9.40E-16&  1.90E-16&  3.40E-16&  1.40E-15&         -&  1.40E-15&         - \\
             & (   2e-17)&(   2e-17)&(   5e-17)&( 1.3e-16)&         -&( 2.8e-17)&( 1.8e-17)&( 2.3e-17)& -4.00E-16&         -&( 6.9e-17)&         - \\
outflow4ap2  &   5.60E-16&  2.00E-15&  5.30E-16&  2.10E-15&         -&  5.80E-16&         -&  1.10E-16&         -&         -&  2.00E-15&         - \\
             & (   2e-17)&( 2.2e-17)&( 2.4e-17)&( 1.2e-16)&         -&( 2.3e-17)&         -&( 1.8e-17)&         -&         -& -2.00E-16&         - \\
     IR6ap1  &          -&  1.10E-15&         -&  9.30E-16&         -&  4.30E-16&         -&         -&         -&         -&         -&         - \\
             &          -&(   3e-17)&         -&( 1.4e-16)&         -&( 3.2e-17)&         -&         -&         -&         -&         -&         - \\
    IR93ap1  &          -&  6.60E-15&         -&  2.70E-15&         -&         -&         -&         -&         -&         -&  5.80E-15&         - \\
             &          -&( 3.5e-17)&         -&(   1e-16)&         -&         -&         -&         -&         -&         -&( 7.4e-17)&         - \\
    IR93ap2  &   4.40E-15&  6.60E-15&  3.90E-15&  3.30E-15&         -&  1.10E-15&         -&         -&  7.60E-15&  5.10E-15&  6.90E-15&  5.50E-15 \\
             & ( 3.2e-17)&( 3.7e-17)&( 9.2e-17)&( 1.4e-16)&         -&( 2.7e-17)&         -&         -&(   8e-17)&( 5.2e-17)&( 7.4e-17)&( 6.1e-17) \\
    IR93ap3  &   1.00E-15&  1.70E-15&         -&  9.00E-16&         -&         -&         -&         -&  2.00E-15&  1.70E-15&  1.90E-15&         - \\
             & ( 2.3e-17)&( 3.6e-17)&         -&( 1.2e-16)&         -&         -&         -&         -&(   8e-17)&( 3.8e-17)&( 8.8e-17)&         - \\
    IR93ap4  &   2.60E-15&  3.70E-15&         -&         -&         -&         -&         -&         -&  4.30E-15&  3.50E-15&  4.70E-15&         - \\
             & ( 3.2e-17)&( 3.6e-17)&         -&         -&         -&         -&         -&         -&( 8.5e-17)&( 5.2e-17)&( 7.4e-17)&         - \\
    IR93ap5  &          -&  1.90E-15&         -&  1.00E-15&         -&         -&         -&         -&         -&         -&         -&         - \\
             &          -&(2.4e-17) &         -&(1.00e-16)&         -&         -&         -&         -&         -&         -&         -&         - \\
    IR93ap6  &          -&  4.10E-16&         -&         -&         -&         -&         -&         -&         -&         -&         -&         - \\
             &          -&(   3e-17)&         -&         -&         -&         -&         -&         -&         -&         -&         -&         - \\  
\enddata
   \tablecomments{Fluxes are in units erg s$^{-1} \persc $\AA$^{-1}$.  Errors are listed on
   the second row for each aperture.  Errors of (0) indicate that the line was detected, but that
   the fluxes should not be trusted because the background was probably oversubtracted.}
\end{deluxetable}\addtocounter{table}{-1}

\begin{deluxetable}{ccccccccccccccc}
  \scriptsize
  \tabletypesize{\tiny}
  \tablewidth{0pt}
    \centering
    \tablecaption{Measured \hh\ line strengths (cont'd)
      \label{tab:nirmeas2}}
    \tablehead{
&2-1 S(1)&2-1 S(3)&3-2 S(3)&3-2 S(4)&4-3 S(5) & [Fe II] & [Fe II] \\
&2.24772&2.07351&2.2014&2.12797&2.20095       & 1.6435  & 1.2567  \\ }
\startdata
outflow1ap1  &   2.00E-15&  1.20E-15&  6.20E-16&  2.60E-16&  7.10E-16 & 4.4e-15    &3.5e-15    \\
             & ( 2.5e-17)&( 3.5e-17)&( 2.2e-17)&( 1.6e-17)&( 1.9e-17) & ( 7.9e-17) &(   4e-17) \\
outflow1ap2  &          -&         -&         -&         -&         - & 6.7e-16    &3.1e-16    \\
             &          -&         -&         -&         -&         - & ( 7.8e-17) &( 3.3e-17) \\
outflow1ap3  &   9.90E-16&         -&  5.70E-16&  2.50E-16&  6.40E-16 & 1.3e-15    &5.7e-16    \\
             & ( 2.6e-17)&         -&(       0)&( 1.2e-17)&(       0) & ( 8.9e-17) &(   4e-17) \\
outflow1ap4  &          -&         -&         -&         -&         - & -          & -         \\
             &          -&         -&         -&         -&         - & -          & -         \\
outflow2ap1  &          -&         -&         -&         -&         - & -          & -         \\
             &          -&         -&         -&         -&         - & -          & -         \\
outflow2ap2  &   6.40E-16&         -&         -&         -&         - & -          & -         \\
             & ( 1.9e-17)&         -&         -&         -&         - & -          & -         \\
outflow2ap3  &          -&         -&         -&         -&         - & -          & -         \\
             &          -&         -&         -&         -&         - & -          & -         \\
outflow4ap1  &   4.30E-16&  4.20E-16&         -&         -&  1.60E-16 & -          & -         \\
             & ( 1.9e-17)& (0)      &         -&         -& (0)       & -          & -         \\
outflow4ap2  &          -&         -&         -&         -&         - & -          & -         \\
             &          -&         -&         -&         -&         - & -          & -         \\
     IR6ap1  &          -&         -&         -&         -&         - & -          & -         \\
             &          -&         -&         -&         -&         - & -          & -         \\
    IR93ap1  &          -&         -&         -&         -&         - & -          & -         \\
             &          -&         -&         -&         -&         - & -          & -         \\
    IR93ap2  &   3.80E-15&         -&  3.10E-15&         -&         - & -          & -         \\
             & ( 2.2e-17)&         -& (0)      &         -&         - & -          & -         \\
    IR93ap3  &          -&         -&         -&         -&         - & -          & -         \\
             &          -&         -&         -&         -&         - & -          & -         \\
    IR93ap4  &          -&         -&         -&         -&         - & -          & -         \\
             &          -&         -&         -&         -&         - & -          & -         \\
    IR93ap5  &          -&         -&         -&         -&         - & -          & -         \\
             &          -&         -&         -&         -&         - & -          & -         \\
    IR93ap6  &          -&         -&         -&         -&         - & -          & -         \\
             &          -&         -&         -&         -&         - & -          & -         \\             
   \enddata
   \tablecomments{Fluxes are in units erg s$^{-1} \persc $\AA$^{-1}$.  Errors are listed on
   the second row for each aperture.  Errors of (0) indicate that the line was detected, but that
   the fluxes should not be trusted because the background was probably oversubtracted.}
\end{deluxetable}
\clearpage

\subsection{Spectroscopic Results: Optical}
\label{sec:dis}

IR 6 and IR 41 (objects 1 and 6 in Figure \ref{fig:outflowsh2}) both show
\ha\ in emission.  IR 41 is close to the reflection nebula in the southeast
portion of \region\ and is probably the reflected star.   The reflection
nebula's spectrum is very similar to IR 41's spectrum at \ha\ in both width
and brightness (see Figure \ref{fig:outflow10_pv}).

\Figure{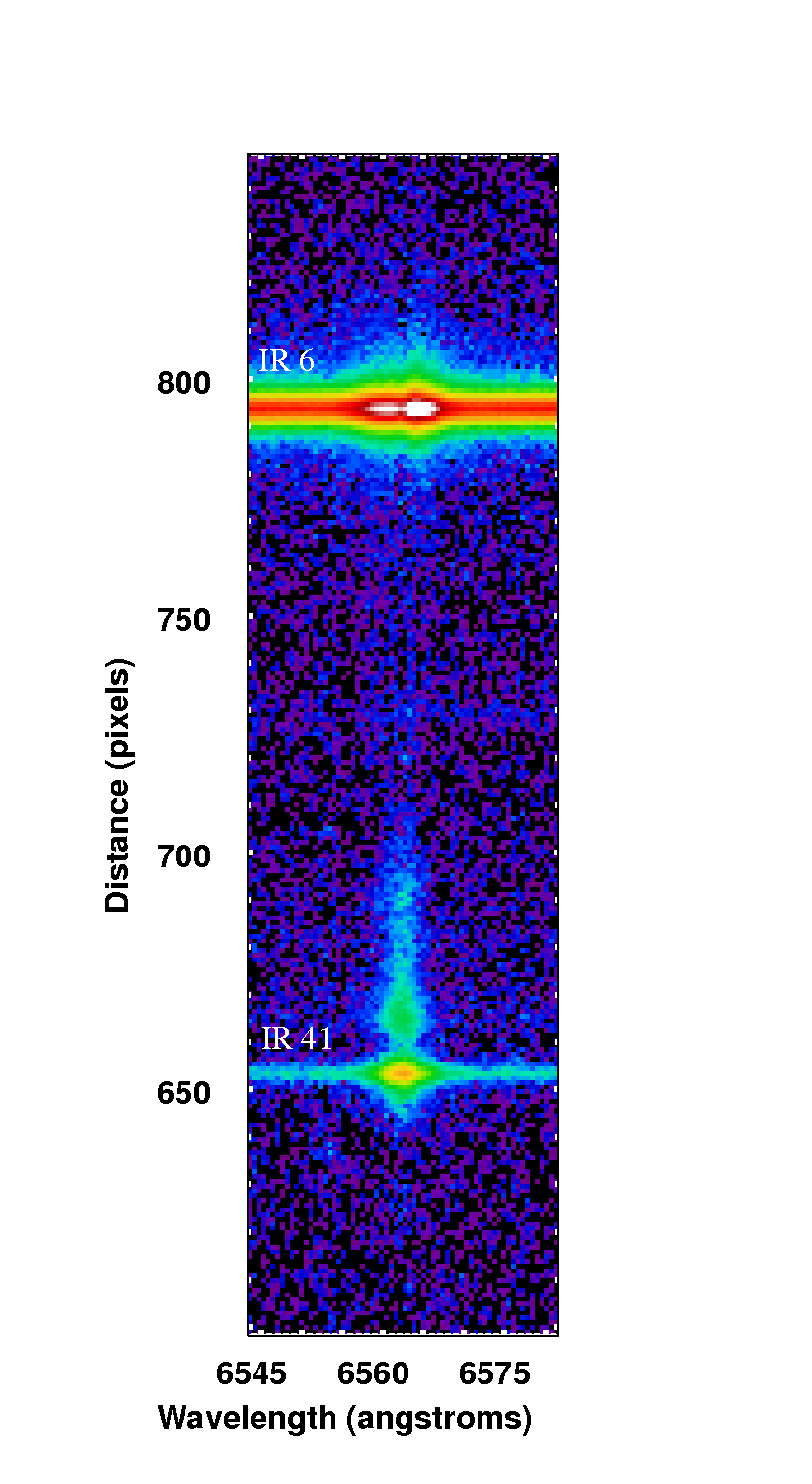}{A position-velocity diagram of IR 6 and 41
including the reflection nebula near IR 41 ($\approx 7.4\msun$).  IR 6 shows a
two-peaked \ha\ emission profile, but is the less massive ($\approx 4.5\msun$) of
the pair.  The separation between the two sources is 55\farcs3, and each pixel
is 0\farcs4.}{fig:outflow10_pv}{0.25}

There are three
components in the \ha\ profile of IR 6: a broad absorption feature seen far
($\sim400\kms$ from the line center) on the wings and two emission peaks.  The
peaks are separated by 190 \kms\ and the blueshifted peak is weaker than the
redshifted (Table \ref{tab:IR6}).  The H$\beta$ profile shows much deeper
absorption and weaker emission but with similar characteristics.   The presence of the \ha\ emission
makes identification of the stellar type from the \ha\ line profile uncertain.
The derived extinction to IR 6 is at least $A_V=7$ from an assumed \ha/\hb\
ratio of 2.87 \citep{agnsquared}.  The \hb\ flux was measured from zero to the
peaks of the emission profile and therefore probably overestimates the
\hb\ flux and underestimates the extinction.

\Table{cccccccc}
{IR 6 Deblended Profiles}
{& \tablenotemark{a} Blue & \tablenotemark{b}Blue & \tablenotemark{a} Red & \tablenotemark{b}Red & Absorption &
Gaussian / & \tablenotemark{b}Absorption \\
&Emission & Wavelength & Emission & Wavelength & & Lorentzian FWHM & Wavelength \\
}
{tab:IR6}
{\ha\ & 4.4\ee{-14} & 6559.79 & 1.3\ee{-13} & 6564.23 & -2.6\ee{-14} & 1.5 / 0.19 & 6563.02 \\
\hb & \tablenotemark{c}2.4\ee{-14} & 4857.68 & \tablenotemark{c}1.8\ee{-14} & 
    4864.28 & \tablenotemark{d} -4.6\ee{-14} & 0.17 / 16.5 & 4861.91 \\}
{
\linebreak
\tablecomments{Measurements are made using a Voigt profile fit in the IRAF {\sc splot} task.}
\tablenotetext{a}{Flux measurements are in units of erg s$^{-1}$ \persc \AA$^{-1}$}
\tablenotetext{b}{Wavelengths are in Geocentric coordinates.  Subtract 0.53\AA\ from \ha\
and 0.39\AA\ from \hb\ to put in LSR coordinates.}
\tablenotetext{c}{\hb\ emission was measured assuming a continuum of zero and therefore represents
an upper limit in the \hb\ emission}
\tablenotetext{d}{\hb\ deblending may contain systematic errors
from a guessed subtraction of the \hb\ emission}
}

\Table{cccccccc}
{Lines observed in the optical spectra}
{Source & \ha\ & \hb\ & [S II] 6716\AA & [S II] 6731\AA & [O I] 6300\AA & [O I] 6363\AA &  [O I] 5577\AA }
{tab:optical}
{ 
Outflow1 ap1                   & 4.3\ee{-16} & - & 5.7\ee{-16} & 6.3\ee{-16} & 5.3\ee{-16} & 1.8\ee{-16} & - \\
                               & 6561.49     & - & 6715.3      & 6729.6      & 6299.7      & 6363.3      & - \\
Outflow1 ap2                   & 4.5\ee{-16} & - & 4.5\ee{-16} & 4.6\ee{-16} & 3.1\ee{-16} & 1.2\ee{-16} & - \\
                               & 6561.22     & - & 6714.9      & 6729.3      & 6299.4      & 6363.2      & - \\
Ambient Medium - slit 1        & 6.7\ee{-17} & 5.3\ee{-18}   & 1.0\ee{-17} & 7.9\ee{-18} & 3.5\ee{-16} & 1.2\ee{-16} & 4.8\ee{-17}   \\ 
                               & 6562.87     & 4861.7        & 6716.7      & 6731.2      & 6300.3      & 6363.8      & 5578.0  \\ 
IR 41 nebula                   & 2.6\ee{-15} & - & - & - & 4.4\ee{-16} & 1.9\ee{-16} & - \\
                               & 6562.85     & - & - & - & 6300.3      & 6363.9      & - \\
IR 41                          & 6.5\ee{-15} & - & - & - & 1.1\ee{-16} & 7\ee{-17}   & - \\
                               & 6562.9      & - & - & - & 6300.0      & 6363.3      & - \\
IR 6                           &  1.76\ee{-13} & \tablenotemark{a} 4.1\ee{-14}    &-&-&-&-   &  \\ 
                               & -             & -                                &-&-&-&-   &  \\ 
}{
\linebreak
\tablecomments{
Wavelengths listed are in \AA\ and are geocentric.  To convert to LSR
velocities, subtract 24.35 \kms.  The ambient medium fluxes represent averages
across the slit.
Fluxes are in erg s$^{-1} \persc $\AA$^{-1}$. 
}
\tablenotetext{a}{\hb\ measurement in IR 6 is an upper limit}
}

\subsection{Radio Interferometry}
\label{sec:vlaresults}
\begin{figure*}[htpb]
  \epsscale{0.75}
  \plotone{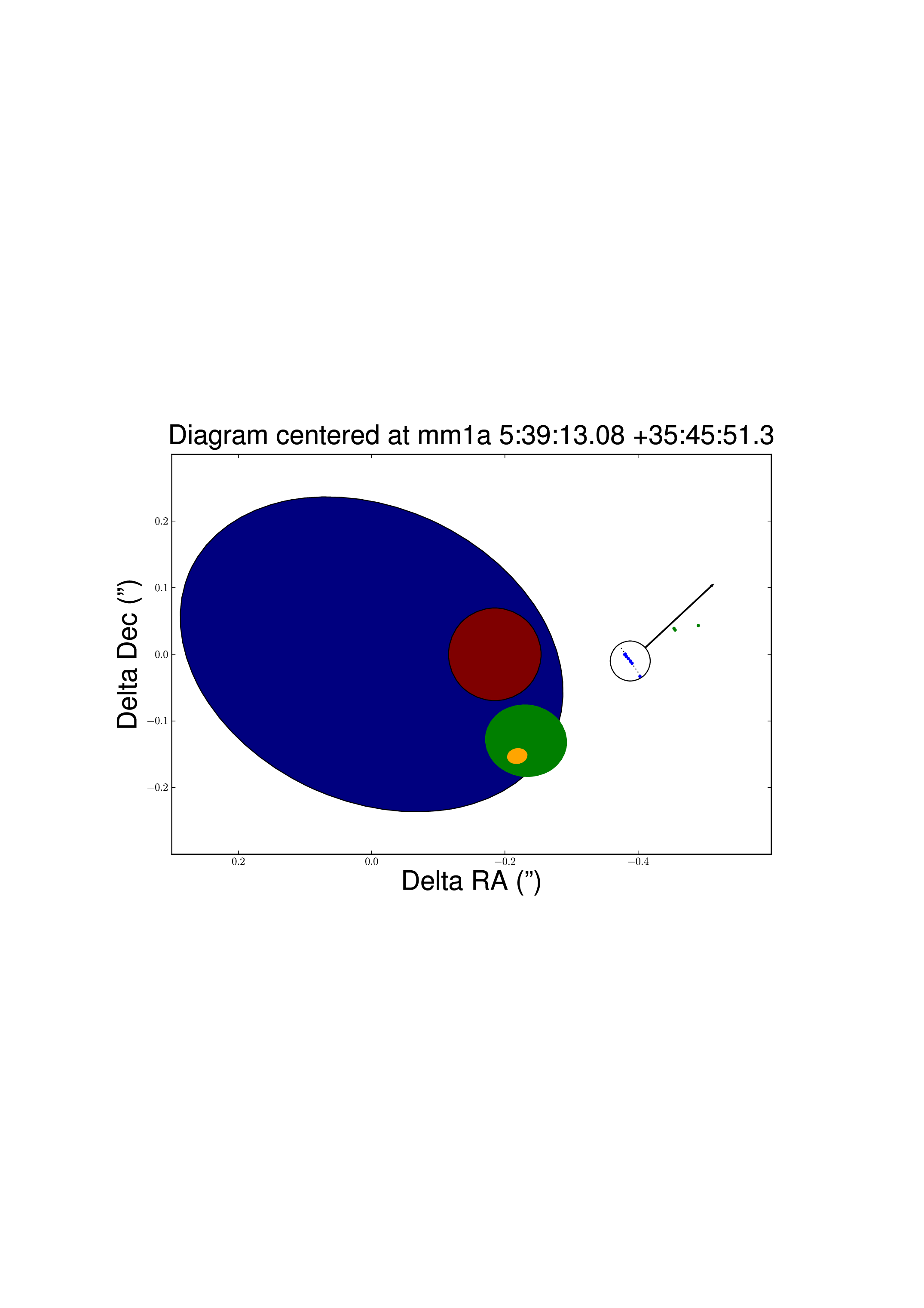}
  \caption{A diagram of the region surrounding mm1a from \citet{beuther2007}.
    The ellipses are centered at the measured source centers and their sizes
    represent the beam sizes of the Plateau de Bure interferometer at 1.2mm
    \citep[blue,][]{beuther2007}, Gemini MICHELLE at 7.9\um\
    \citep[red,][]{Longmore2006}, the VLA at 3.6cm (green), and the VLA at 7mm
    (orange). The maser disk was measured with the European VLBI Network by
    \citet{Minier2000}, so the size and direction of the disk are very well
    constrained.  The black circle is centered on the pointing center of the
    VLBI observation and represents the absolute pointing uncertainty.  The
    arrow pointing in the direction of Outflow 2 traces clumps along the
    outflow back to the mm emission region.  The vector is not to scale -
    Outflow 2 is about 45\arcsec\ long.
\label{fig:mm1adiagram}}
\end{figure*}

A point source was detected in the X, U, K and Q band VLA maps with high
significance at the same location as the X-band point source reported in
\citet{beuther2007}.  Seven-parameter gaussians were fit to each image to measure
the beam sizes and positions and flux densities.  The measurements are listed
in Table \ref{tab:vla}. The locations of the point source and the shape of the
beams from the re-reduced X and Q band images are displayed in figure
\ref{fig:mm1adiagram}. 
A Class II 6.7 GHz methanol maser was detected in \region\ by \citet{Menten1991}.
It was observed with the European VLBI Network (EVN) by \citet{Minier2000} and 
seen to consist of a linear string of maser spots that trace a probable disk in
addition to maser spots scattered around a line perpendicular to the proposed
disk (see Figure \ref{fig:mm1adiagram}).
The VLA source is more than a VLA beam away from the VLBI CH$_3$OH maser disk
identified by \citet{Minier2000}.  It is to the southeast in the opposite
direction of Outflow 2.  Outflow 2 is at position angle -47$\degree$, while the
disk is at PA 25\degree.  The 8$\degree$ difference from being perpendicular is
well within the error associated with determining the angle of the outflow in
this confused region, so the VLBI disk is a strong candidate for the source of Outflow 2.

The astrometric uncertainty in VLA measurements are typically
$\lesssim0.1$\arcsec. Different epochs of high-resolution X-band and Q-band
data confirmed that the pointing accuracy is substantially better than
0.1\arcsec\ in this case.  The VLBI absolute pointing uncertainty is reported
to have an upper limit of $\sim 0.03$\arcsec\ \citep{Minier2000}.  The
separation between the VLA Q-band center and the VLBI disk center is
0.22\arcsec, whereas the separation between the combined X and Q band pointing
centers is only 0.027\arcsec, which can be viewed as a characteristic
uncertainty. This is evidence for at least two distinct massive stars in a
binary separated by $\sim$ 400 AU.  While the statistical significance of the
binary separation is quite high using formal errors, the systematic errors
cannot be constrained nearly as well.  This object is a candidate binary system
but is not yet confirmed.

\begin{deluxetable}{lllll}
    \tablecolumns{7}
    \tablecaption{VLA measurements near \region\ mm1a
  \tabletypesize{\footnotesize}
    \label{tab:vla}}
\footnotesize
\tablehead{ 
\colhead{Frequency } & \colhead{Beam major /} & \colhead{RA (error)}   & 
\colhead{Peak flux }  & \colhead{Map RMS }   \\
\colhead{Observed} & \colhead{minor / PA} & \colhead{Dec (error)} &  \colhead{(error)} & \colhead{(mJy/beam)} \\ }
\startdata
43.3 GHz & 0.022\arcsec\  / 0.029\arcsec\ / -10.4 \degree & 05:39:13.065425 (0.000015) &  1.319 (0.027) & 0.179 \\
                                                       &&35:45:51.14732 (0.00031)    &  \\
22.5 GHz &  1.52\arcsec\  / 1.28\arcsec\  / 232 \degree    & 05:39:13.05521 (0.0029)    &  1.26 (0.04)   & 0.091 \\
&&35:45:51.378 (0.046)        &\\
15.0 GHz & 1.58\arcsec\   / 2.00\arcsec\  / 0 \degree      & 05:39:13.062 (0.005)       &  1.274 (0.065) & 0.124 \\
&&35:45:51.4 (0.1)            &\\
8.4 GHz  &  0.107\arcsec\ / 0.122\arcsec\ / 7.9 \degree  & 05:39:13.064548 (0.000036) &  0.506 (0.003) & 0.015 \\
&&35:45:51.170356 (0.000613)  &\\
\enddata
\tablenotetext{.}{Errors reported here are fit errors.  Absolute flux calibration
errors are negligible for the X-band data but are about equivalent to measurement errors
for the K, and U bands and dominant in the Q band}
\end{deluxetable}

\section{Analysis}

\subsection{Near-Infrared Spectroscopic Extinction Measurements} 
Extinction along a line of sight can be calculated using the 1-0 Q(3) / 1-0 S(1) line ratio. 
\begin{equation}
    A_\lambda = 1.09 \left[ -\textrm{ln}
    \frac{S_\nu(S)/S_\nu(Q)}{A_{ul}(S)\lambda_Q/A_{ul}(Q) \lambda_S} \right] \left[
    \left(\frac{\lambda_S}{\lambda_Q}\right)^{-1.8} -1 \right]^{-1}
\end{equation}
Because they are from the same upper state,
their intensity ratio should be set by their Einstein A values times the
relative energies of the transitions.  
However, as shown by
\citet{luhman1998}, narrow atmospheric absorption lines in the long wavelength
portion of the K band, where the Q branch lines lie, can create a significant
bias.  Because the lines have not been corrected for atmospheric absorption,
the Q branch fluxes should actually be lower limits.  Since the 1-0 S(1)
transition at 2.1818 microns is affected very little by atmospheric absorption,
and the exinction measured is proportional to the Q/S line ratio, the measured
extinction should be a lower limit.

The [Fe II] 1.6435 and 1.2567 \um\ lines were detected in Outflow 1, allowing for
another direct measurement of the extinction.  The measured ratio $FR$ =
1.26\um/1.64\um\ in Outflow 1 was .8, while the true value is at least 1.24 but
may be as high as 1.49 \citep{Smith2006,luhman1998,Giannini2008}.  The
extinction measured from this ratio ranges from $A_V = 4.1$ ($FR=1.24$) to 5.8
($FR=1.49$).  The S(1)/Q(3) ratio uncorrected for telluric absorption is .91,
which yields an extinction lower limit of $A_V = 18.7$, is inconsistent with
the measurement from [Fe II].  The \ha\ detection and \hb\ upper limit give a
lower limit on the extinction of $A_V = 6.6$, which is consistent with both of
the other methods to within the calibration uncertainty.  

It is possible that the two measurements come from unresolved regions with
different levels of extinction, though a factor of at least 3 change over an
area $\sim 100 AU$ far from the millimeter cores seems unlikely.  A strong IR
radiation field could plausibly change the line ratio from the expected
Einstein A value.  The question is not resolved but may be possible to address
with near-IR observations of nearby bright HH flows with more careful
atmospheric calibration.

\subsection{Optical Spectra}
\subsubsection{Stellar Type}
IR 6 is suspected to be the source of the bright \hh\ finger at PA $\approx$
15\arcdeg.  
IR 6 is also
a 24\um\ source and was detected by MSX (designation G173.4956+02.4218).  We
identify this star as a Herbig Ae/Be star. 

\subsubsection{Density and Extinction Measurements}
The spectrum of knot N1 (the bow of Outflow 1) allowed a measurement of
electron density in the shocks from the [S II] 6716/6731  line ratio
.  Densities were determined to be $n=$ 700 \percc\  in the forward lump
and $n=$500 \percc\ in the second lump.  \ha, [N II] 6583, [O I] 6300, and [O
III] 6363 were also detected, but no lines were detected in the blue portion of
the spectrum presumably because of extinction.  The measured velocities from [S
II] are faster than the \hh\ velocity measurements at about $v_{LSR} = -68
\pm 5$ \kms. 

There is also an ambient ionized medium that uniformly fills the slit with a
[S II]-measured density $n_e=120\ \percc$.  Evidently, nearby massive stars are
ionizing the low-density ISM located in front of \region.   This material is
moving at velocity $v_{LSR} = -7 \pm 5$ \kms\ and is extincted by $A_V=1.5$ as
determined from \ha/H$_\beta=2.87$  assuming the gas is at 10$^4$ K.

\subsection{UCHII region measurement}
\begin{figure*}[htpb]
\epsscale{0.75}
\plotone{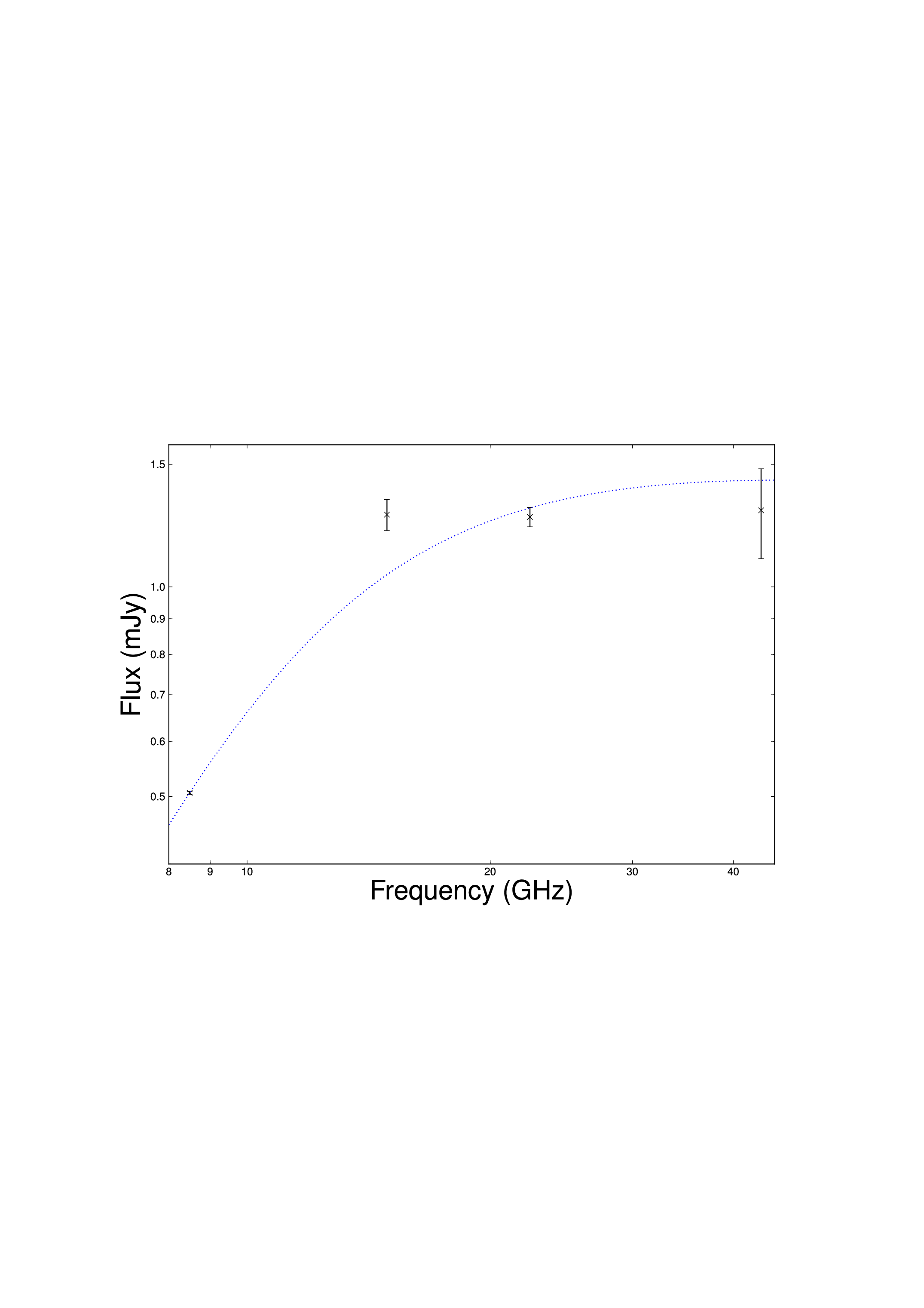}
\plotone{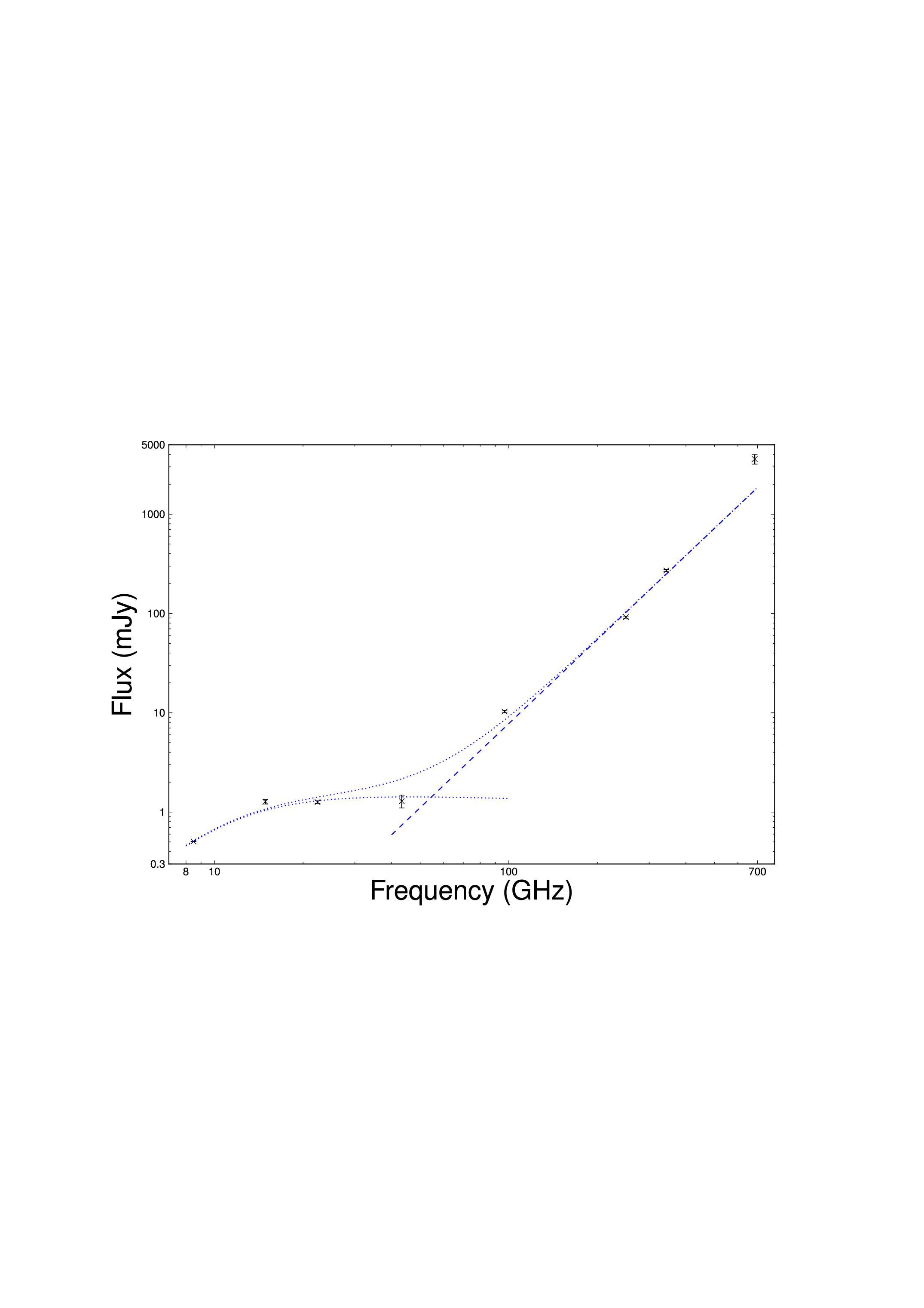}
\caption{(a) The HII region fit to measured X, K, U, and Q band data.  Error bars
represent statistical error in the flux measurement.  The Q band error is
dominated by flux calibration uncertainty (see Table \ref{tab:vlatimes}).  The
measured turnover is at 9.5 GHz. (b) A fit to both the VLA data presented in this paper
and the (sub)mm points from \citet{beuther2007}.  The best fit spectral index for the 
dust emission is $\alpha=2.8$ ($\beta=0.8$), significantly lower than the $\alpha=3.6$
measured by \citet{beuther2007} without the 0.7 mm data point.}
\label{fig:HIIregionfit}
\end{figure*}

A uniform density, ideal HII region will have an intensity curve $I = I_0
( 1 - e^{-\tau_\nu})$ where 
\begin{equation} \tau = 8.235\times10^{-2}
\left(\frac{T_e}{K}\right)^{-1.35} \left(\frac{\nu}{\textrm{GHz}}\right)^{-2.1}
\left(\frac{\textrm{EM}}{\textrm{pc~cm}^{-6}}\right) a(\nu,T)
\end{equation}
following \citet{rohlfs2004} equation 9.35, where $a(\nu,T) \approx 1$ is a
correction factor.   By assuming an excitation temperature $T_{ex} = 8500$K, 
blackbody with a turnover to an optically thin thermal source was fit to the
centimeter SED.  The turnover frequency from this fit is $\tau=$15.5 GHz,
corresponding to an  emission measure $EM=7.4 \times 10^8$ pc cm$^{-6}$.  This
turnover frequency is lower than the $\sim35$ GHz reported by
\citet{beuther2007}.  The turnover is clearly visible in the U, K and Q data
points in figure \ref{fig:HIIregionfit}.  

By assuming the X-band emission is optically thick, a source size can be derived.
\begin{equation}
    \label{eqn:uchiirad}
    2 r = \left[\frac{S_\nu}{2 k_B T_{ex}}  \lambda^2  D^2\right]^{1/2}
\end{equation}
where D is the distance to the source.  Assuming a spherical UCHII region and a
distance of 1.8 kpc, the source has radius $r=$30 AU (for comparison, the
Q band beam minor axis is $\sim$90 AU, so the region could in principle be
resolved by the VLA + Pie Town configuration).  

The measured density is $n=(EM / r)^{1/2} = 2.2\times10^6$ \percc, with a
corresponding emitting mass $M = n \mu m_H 4/3 \pi r^3 = 1.0\times 10^{-6}$
\msun\ using $\mu=1.4$.  Using \citet{kurtz1994} equation 1,

\begin{equation}
  N_{Lyc} = (8.04\times10^{46} s^{-1}) T_e^{-0.85} \left(\frac{r}{pc} \right)^3 n_e^2
\end{equation}

the number of Lyman continuum photons per second required to maintain
ionization is estimated to be  $N_{Lyc} = 5.9\times10^{44}$ 
, a factor of $\sim4$ lower than measured by \citet{beuther2007} and
closer to a B2 ZAMS star ($\sim11\msun$) than B1 using Table 2 of
\citet{panagia1973}.  If the star has not yet reached the main sequence, it
could be significantly more massive \citep{hosokawa2009}, so our stellar mass
estimate is a lower limit.

The gravitational binding radius of a 11 \msun\ star is $r_g = 2 G M / c_s^2 \approx
190$ AU (the HII region is assumed to be supported entirely by thermal
pressure, which provides an upper limit on the binding radius since turbulent
pressure can exceed thermal pressure).  The UCHII region radius of 30 AU is
much smaller, indicating that, under the assumption of spherical symmetry, the
HII region is bound.

\citet{leurini2007} noted that the CH$_3$CN line profile around this source
could be fit with a binary system with separation $<1100$AU and a total mass
of 7-22 \msun .  This is entirely consistent with our picture of a massive binary
system with a 11 \msun\ star in a UCHII region and another high mass star with
a maser disk.

There are no other sources in the \region\ region to a 5$\sigma$ limit of 0.075
mJy in the X-band, which provides the strictest upper limit.  From equation
\ref{eqn:uchiirad}, this corresponds to an optically thick source size of 24
AU.  The maser disk has a spatial extent of around 140 AU, so it is quite unlikely
that either an undetected UCHII region or the observed UCHII are associated with
the maser disk.

Assuming the same turnover point for undetected sources,  
an upper limit is set on $N_{Lyc}$ for undetected sources:
\begin{equation}
  N_{Lyc} = (8.04\times10^{46} s^{-1}) \left(\frac{S_\nu }{  2 k_B T_{ex}^{1.85} }  \lambda_{cm}^2  D_{pc}^2\right) EM
\end{equation}
Our 5$\sigma$ upper limit is $N_{Lyc} = 1.38\ee{44}$ s$^{-1}$, indicating that
any stars present must be a later class than B3, or lower than about 8 \msun .  
For an emission measure as much as 3 times higher, the corresponding stellar
mass would be less than 10 \msun .  It is likely that no other massive
stars have formed in \necluster.

After independently determining the best-fit UCHII model to the VLA data, we
included the PdBI data points from \citet{beuther2007} and fit a power-law to
both data sets.  If the emission measure was allowed to vary, the derived
parameters were $EM=6.3\ee{8}$ and $\beta=0.7$.  However, doing this visibly
worsened the UCHII region fit without significantly improving the power-law
fit, so the fit was repeated holding a fixed emission measure, yielding
$\beta=0.8$ (plotted in Figure \ref{fig:HIIregionfit}b).  This power-law is
much shallower than the $\beta=1.6$ measured by \citet{beuther2007} without
access to the 44 GHz data point, and suggests that there is a significant
population of large grains in source mm1a.  However, we caution that the fits
were performed only accounting for statistical errors, not the significant and
unknown systematic errors that are likely to be present in mm interferometric
data.  The PdBI beams are much larger than the VLA beams, so the larger beams
could be systematically shifted up by including additional emission, which
would reduce $\beta$.  Nonetheless, the new VLA data constrains the UCHII
emission to contribute no more than 10\% of the 3.1mm flux.

\subsection{Mass, Energy, and Momentum estimates from CO}

\subsubsection{Equations}
The column density for CO J=3-2 is estimated using the equation 
\begin{equation}
  \label{eqn:column}
N_{\hh} =
\frac{\hh}{\textrm{CO}}\frac{8\pi\nu^3k_B}{3c^3hB_eA_{ul}}(1-e^{h\nu/k_BT_{ex}})^{-1}
\frac{1}{\eta_{mb}} \int T_A^*(v) dv 
\end{equation}
where $A_{ul}=A_{32}=2.5\times10^{-6}\textrm{s}^{-1}$ and
$A_{21}=6.9\ee{-7}$s$^{-1}$\citep{turner1977}, the rotational constant $B_e =
57.64$, 55.10, and 55.89  GHz for \twelveco, \thirteenco, and \ceighteeno\ respectively,
$\eta_{mb} = .68$, and $T_{ex}$ is assumed to be 20K.  The partition function
is approximated as 
\begin{equation}
    Z=\sum_{J=1}^\infty (2J+1) exp
  \left(\frac{-J(J+1)hB_e}{k_B T_{ex}}\right) \approx \int_0^\infty (2J+1)exp
  \left(\frac{-J(J+1)hB}{k_bT_{ex}}\right) dJ
\end{equation}
which is valid when $T_{ex} >> hB_e/k_B \sim 2.8 $K.
Equation \ref{eqn:column} becomes 
\begin{equation} 
  N_{\hh} = ( 3.27\times10^{18} \persc) \frac{1}{\eta_{mb}} \int T_A^*(v) dv
\end{equation}
where the integrand is in units K \kms.  The mass is then 
\begin{equation}
  M = \mu\ m_{\hh}\ A\ N_{\hh} = 1.42\times10^{-5} A \frac{1}{\eta_{mb}} \int
  T_A^*(v) dv
\end{equation} 
where A is the area in cm$^2$, $\mu=1.4$ is a constant to account for the presence of
helium, and again velocity is in \kms.

\subsubsection{CO J = 2-1 Isotopologue Comparison}
\label{sec:co21}

\citet{Thomas2008} observed C$^{17}$O in the J=2-1 and 3-2 transitions each with a
single pointing using the JCMT centered at J(2000) = 05:39:10.8 +35:45:16 and measured a
column density $N_{\hh}=4.03\times10^{22}$ cm$^{-2}$.  The peak column
density is $1.7\ee{22}$\persc\ in \thirteenco\ and $2.2\ee{22}$ in \ceighteeno\ at J(2000) = 5:39:10.2
+35:45:26, which is reasonably consistent with the C$^{17}$O measurement
considering abundance uncertainties.  The peaks of the integrated
spectra for \ceighteeno\ and \thirteenco\ are coincident, but the \twelveco\
integrated peak is at J(2000) = 5:39:12.6 +35:45:46 (Figure \ref{fig:scuba_co21},
discussed more in section \ref{sec:discussion-outflows}).

Measurements of the column density, mass, momentum, and energy are performed as
in Equation \ref{eqn:column}.  Assuming a \twelveco/\thirteenco\ ratio of 60
\citep{lucas1998} and optically thin \thirteenco, the mean column density
across the region is $N_{\hh} = 4.5\times10^{21}$ \persc.  The resulting total
mass of the central $\sim3$\arcmin\ is about 320 \msun, which is substantially
smaller than the 600 \msun\ measured by \citet{beuther2002} and
\citet{Zinchenko1997}, but it is nearly consistent with 870\um\ and NH$_3$
estimates of 450 and 400 \msun\ from \citet{Mao2004} and is within the
systematic uncertainties of these measurements.  Assuming \ceighteeno\ is
optically thin and the \ceighteeno/\thirteenco\ ratio is 10, the 
column density is 5.2\ee{21} \persc\ and the mass is 360 \msun, which is
consistent with the \thirteenco\ measurements, indicating that optical depth
effects are probably not responsible for the discrepancy with the dust mass
estimate.

\subsubsection{CO Mass and Energy Measurements for Specific Outflows}

Table \ref{tab:comeas} lists measurements of mass and momentum in
apertures shown in figure \ref{fig:cofig}.  Where red and blue masses are
listed, there is an outflow in the red and blue along the line of sight.  Where
only one is listed, an excess to one side of the cloud rest velocity was
detected and assumed to be accelerated gas from a protostellar outflow.  Blue
velocities are integrated from -33 to -21 \kms.  Red velocities are integrated
from -12 \kms\ to 1 \kms.  All masses are computed assuming CO is optically
thin in the outflow, which leads to a lower bound on the mass; \thirteenco\ 2-1 was
measured to have an optical depth of 0.1 in 7 very high velocity outflows in
\citet{choi1993}, so if a relative abundance \twelveco/\thirteenco =
60 is assumed\citep{lucas1998}, masses increase by a factor of 6.

It is not possible to completely distinguish outflowing matter from the ambient
medium.  While the outflowing matter is generally at higher velocities, the
outflow and ambient line profiles are blended.  A uniform selection of high
velocities was applied across the region but this may include some matter from
the cloud, biasing the mass measurements upward.  Outflows in the plane of the
sky and low-velocity components of outflows will be blended with the cloud
profile, which would lead to underestimates of the outflowing mass.  The
momentum measurements, however, should be more robust because they are weighted
by velocity, and higher velocity material is more certainly outflowing.  The
momentum measurements are referenced to the central velocity of \necluster,
$v_{LSR}=-16.0$ \kms.

\section{Discussion}

\subsection{Outflow Mass and Momentum}
\label{sec:discussion-outflows}
\citet{beuther2002} reported a total outflowing mass of 20 \msun\ in \necluster.  We
measure a significantly lower outflow mass of 2 \msun\ under the assumption
that the gas is optically thin, but this assumption is not valid: a lower limit
can be set from the weak \thirteenco\ 2-1 outflow detection  (lower limit
because not all of the outflowing material is detected) on the outflowing mass
of $\sim 4$ \msun .  \citet{choi1993} measure an optical depth of  \thirteenco\
2-1 $\tau \approx 0.1$ in 7 very high velocity outflows.  
Our \thirteenco\ data suggests that the optical depth is somewhat lower,
 $\tau\approx0.07$. The abundance \twelveco/\thirteenco = 60 is used  \citep{lucas1998}  
to derive a total outflowing mass estimate $M\approx25$ \msun .  The total outflowing mass is
therefore $\sim 4\%$ of the total cloud mass, though most
of the outflowing material is coming from \necluster, so as much as 13\% of the
material in \necluster\ may be outflowing.

The most prominent outflow in \region, Outflow 1, is primarily along
the plane of the sky, so the high velocity CO is likely associated
with the other outflows that have significant components along the line of
sight.  As pointed out in \citet{beuther2002}, the integrated and peak CO are 
aligned with the main mm core.  High-velocity \hh\ near the mm cores and the
blueshifted outflows 2 and 4 all suggest that there are many distinct outflows
that together are responsible for the high velocity CO gas.

The offset between the integrated \thirteenco\ peak and \twelveco\ peak in the 
J=2-1 integrated maps, which corresponds with an offset in the peak of the
integrated CO 3-2 map and the peak temperature observed in CO 3-2, suggests
that the gas mass is largely associated with \swcluster, but the outflowing gas
is primarily associated with \necluster.  The integrated and maximum brightness
temperatures in \thirteenco\ and \ceighteeno\ are also centered near \swcluster, 
which rules out optical depth as the cause of this offset.  CO may be depleted
in the dense mm cores, which would help account for the lower mass estimate
from CO isotopologues relative to dust mass and NH$_3$.  Alternately, the gas
temperature in \swcluster\ may be significantly higher than in \necluster\ 
except in the outflows, which are probably warm.  In this case, the outflowing
\twelveco\ enhances the integrated intensity because of its high temperature
and reduced effective optical depth, but it does not set the peak brightness
because of the low filling-factor of the high-temperature gas.

Because the outflows are seen in \hh, which requires shock velocities $\sim30$
\kms\ to be excited \citep{bally2007}, and because the association between the
high-velocity CO and the plane-of-the-sky \hh\ is unclear,  a velocity of 30
\kms\ is used when estimating the dynamical age.
Assuming the outflow is about 0.5 pc long in one direction (e.g. Outflow 1), 
the dynamical age is 1.6\ee{4} years. Outflow 4, which is around 1 pc
long, is also seen at a velocity of -70 \kms\ LSR, or about -50 \kms\ with
respect to the cloud, and therefore has a dynamic age 2\ee{4} years, which is
consistent.

\subsection{Energy Injection / Ejection}
Using an assumed outflow lifetime of $5\times10^3$ years for $v=100\ \kms$ as a
lower limit (because the full extent of the flows is not necessarily observed)
and $1\times10^5$ as an upper limit (for the CO velocities $\sim10\ \kms$ and
the longest $\sim1$ pc flows), mechanical luminosities of the outflows $L=E/t$
are derived.  The summed mechanical luminosity of the outflows is compared to the
turbulent decay luminosity within a 12\arcsec, 1\arcmin, and 5\arcmin\ radius
centered on \necluster\ in Table \ref{tab:turb}.  \setlength\tabcolsep{3pt}  

\Table{ccccccc}{Comparison of turbulent decay and outflow injection}
{Radius (pc) & $t_{turb}$\tablenotemark{a} (yr) & $L_{turb} (\lsun)$ & $L_{outflows}$\tablenotemark{b} $(\lsun) $
& Binding Energy (ergs) \tablenotemark{c} & Outflow Energy (ergs) & Turbulent Energy (ergs) \tablenotemark{d}}
{tab:turb}
{
0.10  & 2\ee{4} &  20  & 0.03-0.6    & 3.4 \ee{46} & 3.5 \ee{44} & 5.0\ee{46}\\
0.52  & 1\ee{5} &  12  & 0.6 - 9.4   & 5.9 \ee{46} & 5.9 \ee{45} & 1.5\ee{47}\\
2.62  & 5\ee{5} &  2.3 & 1-22        & 1.2 \ee{46} & 1.4 \ee{46} & 1.5\ee{47}\\
}
{
\tablenotetext{a}{Masses are assumed to be 600 \msun\ for the 1\arcmin\ and 5\arcmin\ 
apertures, and 200\msun\ for the 12\arcsec\ aperture.}
\tablenotetext{b}{Outflow luminosities are given as a range with a lower limit
$L=E_{out} / 10^5 \textrm{yr}$ and upper limit $L=E_{out}/ 5\times10^3 \textrm{yr}$, 
where $E_{out}$ is from Table \ref{tab:comeas} multiplied by 6 to correct
for outflow opacity.  }
\tablenotetext{c}{Binding energy is the order-of-magnitude estimate GM$^2$/R}
\tablenotetext{d}{Turbulent energy is computed using the measured 5 \kms\ line
width as the turbulent velocity.}
}

The rate of turbulent decay can be estimated from the crossing-time of the
region,  $L / v$, where $L$ is the length scale and $v$ is the the typical
turbulent velocity.  On the largest ($\sim$ few pc) scales, the mechanical
luminosity from high-velocity outflowing material is approximately capable of
balancing turbulent decay and upholding the cloud against collapse.  However,
at the size scales of the \necluster\ clump ($\sim 0.1$ pc), turbulent decay
occurs on more than an order of magnitude faster timescales than outflow energy
injection.  On the smallest scales, outflow energy can be lost from the cluster
through collimated outflows, though wide-angle flows and wrapped up magnetic
fields will not propagate outside of the core region. Once collimated flows
impact the local interstellar medium in a bow shock, their energy and momentum
are distributed more isotropically and again contribute to turbulence.  The
imbalance on a small size scale is consistent with the observed infall
signature (Figure \ref{fig:co21_all3}) in the inner 12\arcsec\ around
\necluster\ and the lack of a similar profile elsewhere.

\subsection{Comparison to other clumps}
The classification scheme laid out in \citet{klein2005} is used to identify
\necluster\ as a Protocluster and \swcluster\ as a Young Cluster.  \citet{maury2009}
performed a similar analysis of the Early Protocluster NGC 2264-C.  They also
found that the outflow mechanical luminosity could provide the majority of the
turbulent energy $L_{turb}\sim1.2 \lsun$ within the protocluster in a radius of
0.7 pc with a mass 2300\msun .  \citet{williams2003} performed an outflow
study of the OMC 2/3 region with radius 1.2 pc and mass 1100 \msun, which is also an Early
Protocluster, and concluded that $L_{turb} \sim L_{flow} \sim 1.3 \lsun$.
While all three regions have nearly the same turbulent decay luminosities and
outflow mechanical luminosities, \necluster\ in \region\ is significantly more
compact and lower mass than the Early Clusters, and is the only one of the three
that contains signatures of massive star formation.

\subsection{Surrounding Regions}
\label{sec:surroundings}

About 8\arcmin\ to the southeast of \region\ is another embedded star forming
region, G173.58+2.45.  Interferometric and stellar population studies have been
performed by \citet{shepherd1996} and \citet{shepherd2002}.  The bipolar
outflow detected in their interferometric maps is also cleanly resolved in our
figure \ref{fig:cofig}.  In our wide-field \hh\ maps, there is a complex of
outflows similar to that of \region, but fainter.

The large HII region Sharpless 231 to the northeast can be seen in the \ha\ 
image (figure \ref{fig:overview_ha}).  The expanding HII region is pushing against
the molecular ridge that includes \region\ and accelerating the CO gas in the
blue direction (e.g. the northern blueshifted clumps in figures \ref{fig:cofig}
and \ref{fig:HA_with_CO}).  It can be seen from the IRAC 8\um\ data 
that UV radiation from the HII region reaches to
the \region\ clusters.  The expanding HII region's pressure on the molecular
ridge may be responsible for triggering the collapse of \region\ and G173.  The
size gradient from S232 ($\sim 30\arcmin$\ across) to S231 ($\sim 10\arcmin$)
to S233 ($\sim 2-3\arcmin$) is suggestive of an age gradient assuming uniform
HII region expansion velocities and a common distance.  Investigation of this
hypothesis will require detailed stellar population studies in the HII regions
with proper regard for eliminating foreground and background sources.

\subsection{Massive Star Binary}
Our identification of a probable massive star binary with an associated outflow
contributes to a very small sample of known maser disks with \hh\ emission
perpendicular to the disk.  \citet{debuizer2003} observed 28 methanol maser
sources with linear distributions of maser spots in the \hh\ 2.12 \um\ line,
and he identified only 2 sources with \hh\ emission perpendicular to the maser
lines.  None of the outflows identified in his survey were as collimated as
Outflow 2, so the methanol disk / outflow combination presented here may be the
most convincing association of a massive protostellar disk with a collimated
outflow.

The association of a massive star with an UCHII region and a methanol maser
disk and the very small size of the UCHII region both suggest that the massive
stellar system is very young.  \citet{walsh1998} suggested that the development
of a UCHII region leads to the destruction of maser emission regions.  Their
conclusion is consistent with our interpretation of mm1a as a binary system.

\section{Summary \& Conclusion}

We have presented a multiwavelength study of the \region\ star forming region.
\region\ contains an embedded cluster of massive stars and is surrounded by
outflows.  The outflows were linked to probable sources and determined that at
least one outflow is probably associated with a massive ($\sim 10 \msun$) star.
Added kinematic information and a wide field view of the infrared outflows has
been used to develop a more complete picture of the region.

\begin{itemize}
  \item \necluster\ is a Protocluster and \swcluster\ is a Young Cluster
  \item Energy injection on the scales of \region\ can maintain turbulence, but on
    the small scales of the \necluster\ protocluster, is inadequate by $\sim 2$ orders
    of magnitude.  \necluster\ is collapsing.
  \item there are 11 candidate outflows, 7 of which have candidate counterflows, in the 
    \region\ complex
  \item there is a probable massive binary with one member of mass 12 \msun\ in
    mm1a, and the other which is the source of Outflow 2
  \item there are at least two moderate-mass ($\sim$5\msun) young stars in \region\ 
\end{itemize}

We have identified additional middle- and high-mass young stars with outflows, and 
presented a case for a high-mass binary system within the millimeter core mm1a.

\section{Acknowledgements}
We would like to thank Vincent Minier for providing us with the positions of
the VLBI maser spots and Steve Myers and George Moellenbrock for their assistance with
VLA data reduction.

We would also like to thank Cara Battersby, Devin Silvia, Mike Shull, and Jeremy
Darling for helpful comments on early drafts. 

This work made use of SAOIMAGE DS9 (\url{http://hea-www.harvard.edu/RD/ds9/}),
IRAF (\url{http://iraf.net/}, scipy (\url{http://www.scipy.org}), and APLpy
(\url{http://aplpy.sourceforge.net/}).

\end{document}